\documentclass[final,3p,times,twocolumn]{elsarticle}

\usepackage{amssymb}
\usepackage{amsmath}
\usepackage{graphicx}
\usepackage{xr}
\usepackage{mwe}
\usepackage[dvipsnames]{xcolor}
\definecolor{cyanish}{RGB}{200,250,250}
\definecolor{yellowish}{RGB}{250,250,200}
\definecolor{pinkish}{RGB}{250,212,223}
\usepackage{hyperref}
\usepackage{ulem}
\usepackage{wasysym}
\usepackage{rotating}
\usepackage{colortbl}
\usepackage{natbib}
\usepackage{multirow}

\journal{Acta materialia}

\begin{document}

\begin{frontmatter}



\title{Morphological Transition: From Meanders to Mound Structures} 

\author[PAN]{Marta A. Chabowska }
\ead{galicka@ifpan.edu.pl}
\author[BAS]{Hristina Popova }
\author[PAN]{Magdalena A. Za{\l}uska-Kotur } 

\affiliation[PAN]{organization={Institute of Physics, Polish Academy of Sciences},
            addressline={Al. Lotnikow 32/46},
            city={Warsaw},
            country={Poland}}

\affiliation[BAS]{organization={Institute of Physical Chemistry, Bulgarian Academy of Sciences},
            addressline={Acad. G. Bonchev str., 1113},
            city={Sofia},
            country={Bulgaria}}

\begin{abstract}
Mound formation on flat and miscut crystal surfaces exhibits distinct growth behaviors. While mound structures are the predominant feature on flat surfaces, miscut surfaces display a smooth transition from meandered patterns to three-dimensional mounds, depending on both internal and external conditions. We investigate this morphological evolution-from meander-like surface patterns to faceted pyramidal structures-using a Vicinal Cellular Automaton modeling framework.
The transition is shown to be governed by the competition between the Ehrlich-Schwoebel barrier and adatom mobility on terraces. Under moderate barrier strengths and sufficiently high terrace diffusivity,
the system demonstrates a reversible transition from mounded configurations to regular step meandered patterns. This reveals a complex interplay between kinetic barriers and mass transport.
Our simulations cover a wide range of growth conditions, including variations in deposition flux, surface diffusion rates, temperature, and miscut angle. By applying the height-height correlation function,
we calculate the correlation lengths along and across the steps and analyze their scaling behavior.
These results offer insight into the continuum pathways that connect distinct classes of surface structures and provide a unified framework for describing pattern evolution across different crystal growth regimes.
\end{abstract}



\begin{keyword}
Crystal growth \sep surface patterns \sep mounding \sep step meandering \sep computer simulations \sep cellular automata \sep time scaling


\end{keyword}

\end{frontmatter}
\section{Introduction}

The spontaneous formation of three-dimensional (3D) surface patterns during epitaxial crystal growth is a subject of longstanding interest, both from a fundamental and technological perspective. Among them, pyramidal mound structures have been observed across a range of material systems \cite{Damilano-JAP,Chou-ASS,Wu-JCG,Gocalinska-ASS,Pandey-Vac,Uesugi,Krug,Teisseyre-CGD}, and are known to emerge via different  mechanisms, such  as   kinetic instabilities or defect-driven growth processes. Their presence may be beneficial, as in the case of templated self-assembly and nanostructure nucleation, or detrimental when flatness and surface uniformity are desired for device applications.

Understanding how such 3D morphologies develop requires a detailed analysis of the interplay between atomic-scale kinetics and macroscopic growth conditions. Surface orientation, step structure, diffusion anisotropy, and kinetic barriers all contribute to the complexity of pattern formation. On flat crystal surfaces, mound formation often dominates the growth landscape due to the Ehrlich-Schwoebel (ES) barrier, which suppresses downward adatom motion at step edges and promotes vertical mass transport. In contrast, vicinal (miscut) surfaces-characterized by a regular array of atomic steps-support lateral instabilities that can lead to step meandering. These meandered patterns may stabilize into periodic undulations or transition into 3D faceted structures depending on the specific conditions.

For years, scientists have debated the origins of mound formation. In the literature, this topic is approached either analytically \cite{Krug,Siegert,Dasgupta,Rost,Bryan} or atomistically \cite{Krug,Dasgupta,Rost,Evans,Bartelt,Tejedor,Leal,Schneider,Larsson,dasSarma,Munko}. One of the first papers on this subject, \cite{Siegert}, showed how nonlinear instabilities lead to self-organized mound structures. The authors have demonstrated the slope selection and coarsening mechanisms in epitaxial growth through the use of continuum equations. Similarly, Chakrabarti and Dasgupta \cite{Dasgupta} proposed a nonlinear instability mechanism that drives mound formation and coarsening. This mechanism offers an alternative to ES-barrier-driven models. Conversely, Rost et al. \cite{Rost} extended the instability theory to vicinal surfaces, where pre-existing steps along with the existence of step ES barrier change the dynamics. They emphasized the competition between smooth step-flow growth and unstable mound formation. Many other publications \cite{Bartelt,Tejedor,Leal,Schneider,Larsson} have considered the same mechanism to be responsible for mound formation. Punyindu Chatraphorn et al. \cite{dasSarma} proposed yet another approach, studying models of networks with limited mobility. They demonstrated that mounds can form even in the absence of explicit ES barriers.
Two works, \cite{Krug,Evans}, stand out among the literature. These works form the backbone of the field by synthesizing kinetic theory, scaling laws, instabilities, and experimental evidence, as well as emphasizing stochastic effects, kinetic roughening, and step-edge barriers. These studies serve as a benchmark references, tying together atomistic processes and continuum descriptions. Today, the consensus is that multiple mechanisms can interact depending on the growth conditions.

A very interesting perspective that is highly relevant to our study is presented by Rost et al. \cite{Rost} and by Bryan et al. \cite{Bryan}. In the first paper, the authors used continuum models and kinetic Monte Carlo simulations to analyze the instability of step flow during molecular beam epitaxy. They observed transitions from ripples to 3D structures during their studies. In the second paper, the authors used a model based on the Burton, Cabrera, and Frank theory to understand how surface kinetics depend on vapor supersaturation and substrate misorientation angle. The study demonstrated a transition from 3D structures through meanders to step bunches, which depends on the surface misorientation.

Compared to previous studies, the present work focuses on elucidating the transition between two characteristic surface morphologies:
regular meandered step patterns and fully developed faceted pyramidal mounds. Such a transition reflects a shift in the dominant physical processes - between terrace diffusion and asymmetries of interlayer kinetic  - and is a hallmark of the complex dynamical behavior of growing surfaces. While meandered steps arise from lateral instabilities under moderate anisotropy and diffusion rates \cite{Chabowska-PRB}, pyramidal mounds are typically linked to stronger Ehrlich-Schwoebel barriers and reduced adatom mobility. In this work, we employ a novel method to investigate the transition between these two growth regimes.

To explore the transition from meanders to mounds, we employ a vicinal Cellular Automaton (VicCA) modeling framework, which captures the key kinetic features of step flow growth, including adatom diffusion, attachment kinetics, and step-edge barriers. By tuning the ES barrier and surface diffusion parameters, we investigate the morphological evolution of surfaces under a wide range of conditions. We demonstrate that the competition between downward hopping suppression (controlled by the height of the ES barrier) and adatom mobility on terraces governs the observed surface morphology. Crucially, we find that the transition is reversible: high diffusion rates are able to restore meandered patterns even after mound formation has occurred. This suggests that the system resides near a dynamic equilibrium point, where small changes in kinetic parameters produce qualitatively different morphologies.

To quantitatively characterize the transition, we analyze the evolution of height-height correlation functions, surface roughness, and characteristic lateral and vertical length scales. These measures allow us to distinguish between disordered mound growth, regular meandering, and highly ordered faceted pyramids. We further discuss the scaling behavior of these quantities in relation to diffusion rates and step-edge barriers.

Overall, this study offers a unified framework for understanding the morphological evolution of both flat and vicinal surfaces during epitaxial growth. By linking meander formation and mound growth within a common kinetic picture, we provide insight into how surface structures can be controlled by tuning the growth parameters. These results are relevant for the rational design of surface templates and the development of nanostructured materials via bottom-up fabrication techniques.

\section{The model}

The process of surface dynamics is analyzed using a (2~+~1)D vicinal Cellular Automaton model. This model simulates the growth of a vicinal crystal surface through a combination of Cellular Automaton and Monte Carlo techniques, which together govern the evolution of the surface \cite{Chabowska-PRB,MZK-crystals,Chabowska-ACS,Chabowska-Vac,Chabowska-JCG,REDKOV2025}. The model is divided into two interrelated components, enabling independent control of each of them and enhancing both the efficiency of simulations and the model's adaptability.

\begin{figure*}[hbt]
 \centering
a)\includegraphics[width=0.3\textwidth]{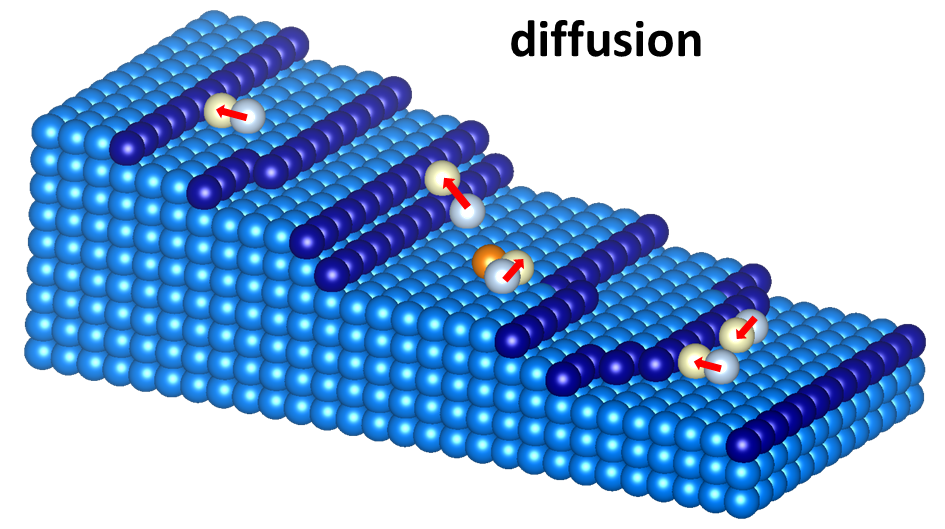}
b)\includegraphics[width=0.3\textwidth]{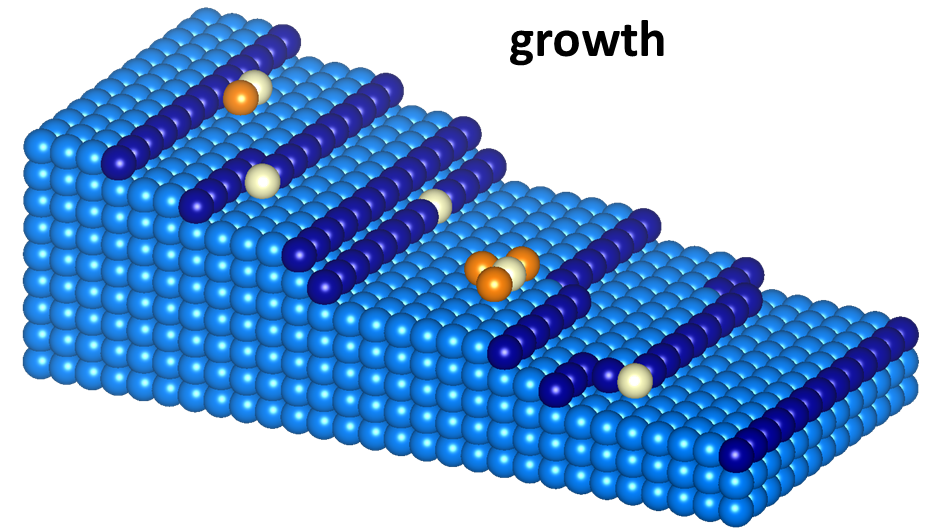}
c)\includegraphics[width=0.3\textwidth]{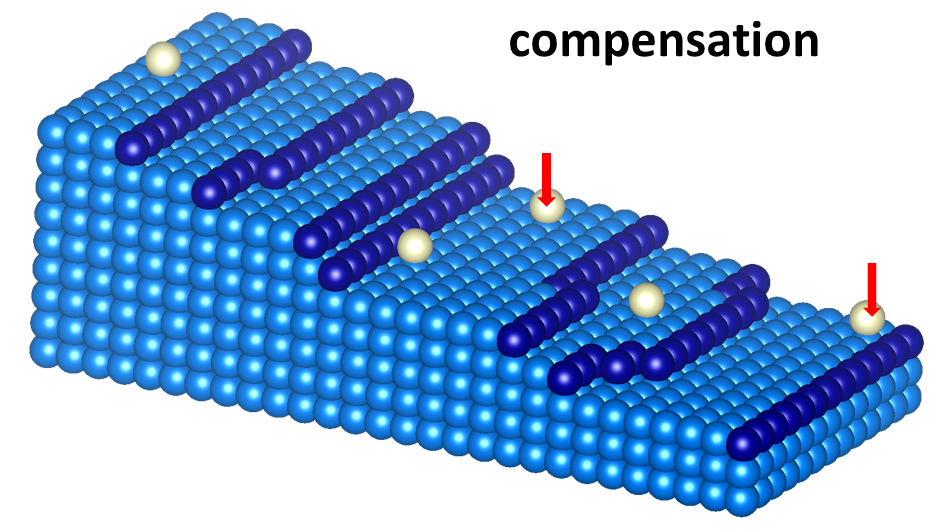}
\caption{A single time step of the simulation procedure consists of the following stages: (a) the diffusion process, (b) the growth update (illustrating different scenarios of adatom incorporation), and (c) the adatom concentration compensation. For the system shown, the number of adatoms must be adjusted to 5 in the last stage of the time step to keep the initial concentration of adatoms constant. In (a) blue balls represent the positions of adatoms before diffusion, yellow - the positions of adatoms after the diffusion step; In (b) yellow balls represent particles freshly incorporated into the crystal,  orange -  represent additional particles whose presence is required for incorporation of the adatom into the crystal; In (c) yellow balls represent the positions of adatoms after compensation.}
\label{fig:procedure}
\end{figure*}

The system comprises of two main parts: the bulk crystal and the surface layer populated by diffusing adatoms, which continuously exchange particles during growth. In real materials, these two regions consist of identical atoms capable of moving and rearranging over time. However, during the growth process, atomic movement within the bulk is limited due to strong interatomic bonding and spatial constraints imposed by newly arriving atoms.
Crystal growth fundamentally involves the integration of atoms into the bulk structure. Prior to incorporation, atoms are supplied from the surroundings and initially adsorbed randomly onto the crystal surface. These adsorbed atoms then diffuse along the surface (Fig.~\ref{fig:procedure}a) in search of energetically favorable sites. This mobile population of atoms forms what is referred to as the surface cloud. Once an atom reaches a particularly stable position, the probability of its desorption becomes negligible.

The crystal component of the model is constructed on a square lattice. The results presented below are based on simulations initiated from a flat or vicinal surface, the latter composed of monoatomic steps descending from left to right, with an initial terrace width denoted by $l_0$. Periodic boundary conditions are applied along the step edges, while helical periodic boundary conditions are used across the steps to preserve the relative height differences between terraces.

To ensure efficient and fast computations, crystal growth is modeled using Cellular Automaton (CA) rules, which enable parallel processing. At each time step, the CA rules determine which atoms are incorporated into the crystal lattice based on the current configuration of atoms and step positions on the surface.
In this model, the most stable incorporation sites are step voids with three neighboring bonds and kinks with two bonds. These specific sites play a crucial role in the vicinal automaton component of the VicCA model. The model assumes that such voids and kinks are filled unconditionally. In contrast, atoms located in less stable positions-such as those with only a single bond to the step or entirely isolated on the terrace-require additional conditions to be met before incorporation is permitted. Consequently, adatoms are allowed to diffuse on top of the terrace or along the step edge until the criteria for stable incorporation are satisfied, effectively enabling control over the step stiffness.

Various methods can be employed to regulate the attachment frequency, such as assigning a low probability of incorporation. Several alternatives were tested, all producing comparable results. Ultimately, we adopted an approach that requires the presence of at least one neighboring particle for successful incorporation (see Fig.~\ref{fig:procedure}b). This condition ensures that an adatom is incorporated into a straight step only if another adatom from the diffusing cloud is adjacent to it. The neighboring atom provides an additional bond, thereby stabilizing the attachment and justifying the incorporation process.
The process can be viewed as nucleation occurring at the step edge, characterized by a critical nucleus size of two  atoms.

The model also allows for independent control of the nucleation process. When nucleation on  top  of  the  terrace  is suppressed, steps advance regularly, exhibiting bunching or meandering behavior, but true three-dimensional growth does not occur. When nucleation is enabled, a critical nucleus size can be defined, adjusting the difficulty of nucleation. In the simulations presented here, a critical nucleus size of four was chosen. According to the CA rules, if an adatom is surrounded by three other adatoms, it becomes incorporated into the crystal, forming a stable nucleus for subsequent growth. This mechanism enables nucleation and thus the formation of three-dimensional nanostructures. Examples of various configurations leading to adatom incorporation are illustrated in Fig.~\ref{fig:procedure}b.

\begin{figure*}[hbt]
\centering
a)\includegraphics[width=0.4\textwidth]{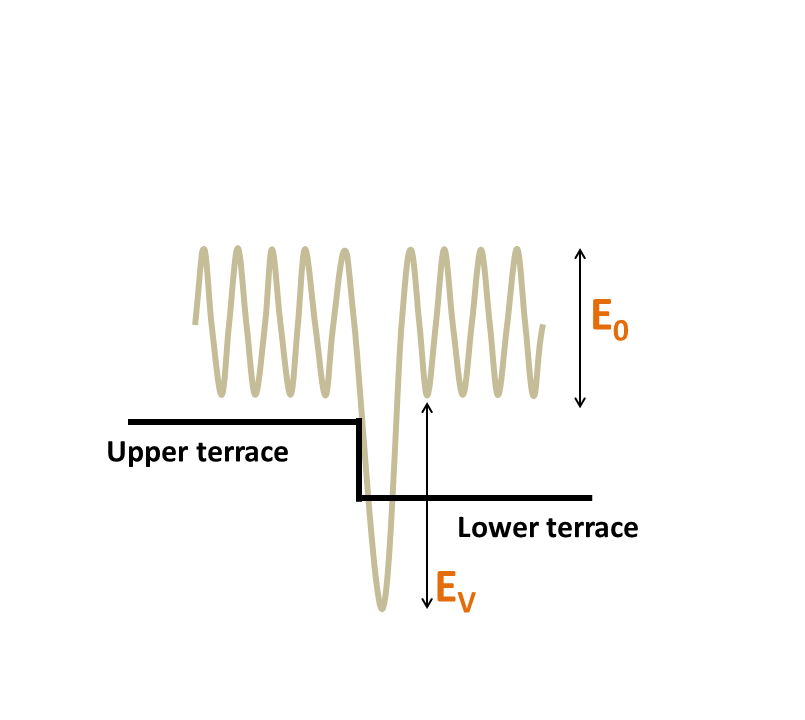}
b)\includegraphics[width=0.4\textwidth]{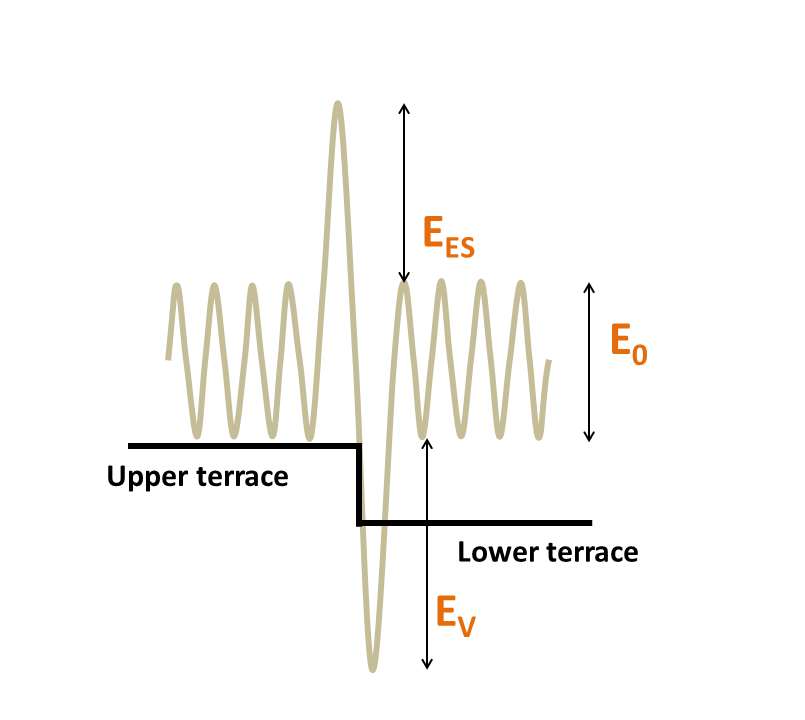}
\caption{Visualization of the potential landscape, including the diffusion barrier $E_0$ and highlighting the potential well at the bottom of the step. Side view of the initial surface with a) only the potential well $E_V$ and b) the additional Ehrlich--Schwoebel barrier $E_{ES}$.}
\label{fig:potential}
\end{figure*}

A key component of the model is adatom diffusion, which acts as the fundamental driving force behind surface pattern formation. This process is strongly influenced by the underlying energy potential landscape experienced by the atoms. Drawing on insights from ab initio studies \cite{Ohka-CGD, Akiyama-JCG20, Akiyama-JCG21, Akiyama-JJAP}, we postulate the existence of a potential well located at the bottom of each step. In our simulations, the depth of this well is kept constant and is denoted by $E_V$ (see Fig.~\ref{fig:potential}a). As demonstrated in \cite{Chabowska-PRB}, both the presence and depth of this potential well play a critical role in triggering and shaping the step meandering process.
The overall shape of the surface potential is further modulated by factors such as dangling bonds, surface reconstruction, and additional energy barriers. Particularly relevant to the development of step meanders is the ES barrier, located at the top of the step (Fig.~\ref{fig:potential}b), which hinders adatom crossing the step. The height of this barrier is denoted by $E_{ES}$.
It is important to note that the shape of the surface potential-as well as the positions of the potential well $E_V$ and ES barrier $E_{ES}$-depend on the positions of the steps and shift accordingly with step displacement. As a result, the potential landscape and step morphology become intricately linked, often evolving in a strongly coupled manner.

In the model, adatoms diffuse independently. On flat terraces, jumps occur with uniform probability, except near step edges where local potentials alter the dynamics. The diffusion rate and the degree of step permeability are regulated by the number of diffusion attempts each adatom makes per unit time, denoted by $n_{DS}$.
Each adatom attempts to perform $n_{DS}$  jumps, with each movement requiring the adatom to overcome an energy barrier between two  consecutive potential  wells,  and  this energy  is  equal  to $ E_0$, referred to as a diffusion barrier.
In the model studied below, all jumps along terrace, away from steps happen with the same probability given by:
\begin{equation}
P_0= e^{-\beta E_0},
\end{equation}
where $\beta=1/(k_BT)$, $T$ is the temperature, and $k_B$ is the Boltzmann constant.
The  diffusion  constant  can  be  expressed  as $D= a^2 \nu n_{DS}P_0$, where $a$ is the lattice constant, used as a unit of length in our simulations, and $\nu$ is the attempt frequency, usually of order $\nu=10^{12}- 10^{13}~Hz$.

During the simulation, we assume that the fastest processes happens with probability equal to 1, which means that the duration of all processes is normalized by the value of $P_0$.
For sites close to steps we have:
\begin{equation}
\frac{P_V}{P_0} = \left\{\begin{array}{rcl}
&e^{-\beta E_V} & \ \text{out}\\
&1 & \ \text{in}
\end{array} \right.
\end{equation}
for jumps out and in the well accordingly,
whereas in the presence of an ES barrier, this probability of a jump across the steps is set to
\begin{equation}
\frac{P_{ES}}{P_0} = \left\{\begin{array}{rcl}
& e^{-(\beta E_{ES}+\beta E_V)}& \ \text{out}\\
& e^{-\beta E_{ES}} & \ \text{in.}
\end{array} \right.
\end{equation}
As in traditional Monte Carlo simulations, adatoms that jump, as well as the jump direction, are selected randomly to prevent artificial correlations.

By appropriately combining diffusion mechanisms with CA rules, various aspects of surface dynamics can be systematically investigated. Each simulation step concludes with replenishing the adatom concentration to a fixed value $c_0$, thereby maintaining a constant external flux of incoming atoms (see Fig.~\ref{fig:procedure}c).

Unlike conventional Monte Carlo simulations, the VicCA model permits independent control over all parameters. In standard MC simulations, jump probabilities and attachment rates are inherently coupled to interaction strengths. In contrast, the VicCA model decouples step attachment from kink attachment and naturally incorporates additional barriers into the potential landscape. Rather than deriving parameters from interactions among multiple atomic neighbors, VicCA allows these parameters to be specified independently on a site-by-site basis.

However, the calibration of time and temperature scales demands careful attention, as these depend on several factors, including the diffusion barrier height $E_0$, the number of diffusional steps $n_{DS}$, the attempt frequency $\nu$, and the external flux $c_0$. The real-time duration of each VicCA simulation step is given by $\tau=(\nu n_{DS} \exp(-\beta E_0))^{-1}$, while the ratio of diffusion to external particle flux is expressed as $D/F = a^2 n_{DS} / c_0$. Temperature enters explicitly into the jump probabilities, thereby becoming an integral part of the overall timescale.  Consequently, temperature indirectly influences the external flux through its effect on the timescale. Basing on the expression  for diffusion coefficient $D=a^2  \nu n_{DS} \exp(-\beta E_0)$, and because we realize all jumps along terrace with the same rate 1, we can relate temperature difference to $n_{DS}$ as $\Delta \beta=\ln(n_{DS}^{-1})/E_0$. The larger $n_{DS}$ is, the lower $\beta$ value and larger temperature $T$.
Additionally, it is important to note that in all our simulations, potential energies $ E_V$ and $ E_{ES}$ are measured in units of thermal energy $k_BT$.

\section{Mound  formation}
\begin{figure*}[hbt]
 \centering
\includegraphics[width=0.8\textwidth]{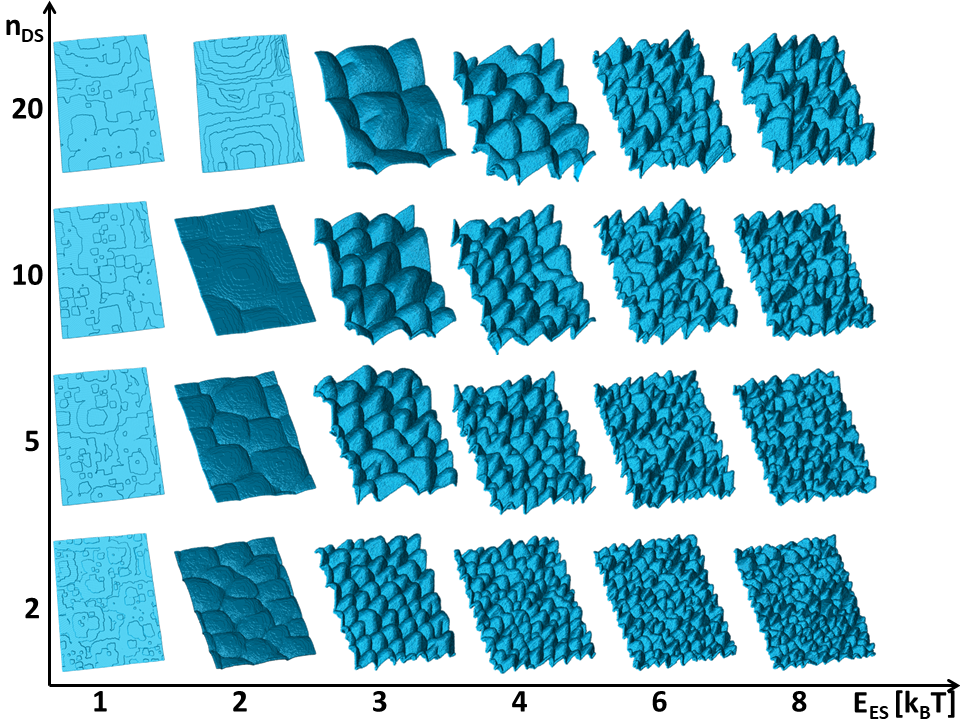}
\caption{Diagram of different surface patterns presented as a function of number of diffusional jumps $n_{DS}$ dependent on the height of  ES  barrier $E_{ES}$ for flat surface, $E_V = 1.0 k_BT$ and initial particle concentration $c_0=0.01$. System size $300$ x $400$ (in units of the lattice constant) and number of simulation steps $t=10^6$.}
\label{ev-vs-es-L300}
\end{figure*}
\begin{figure*}[hbt]
 \centering
\includegraphics[width=0.8\textwidth]{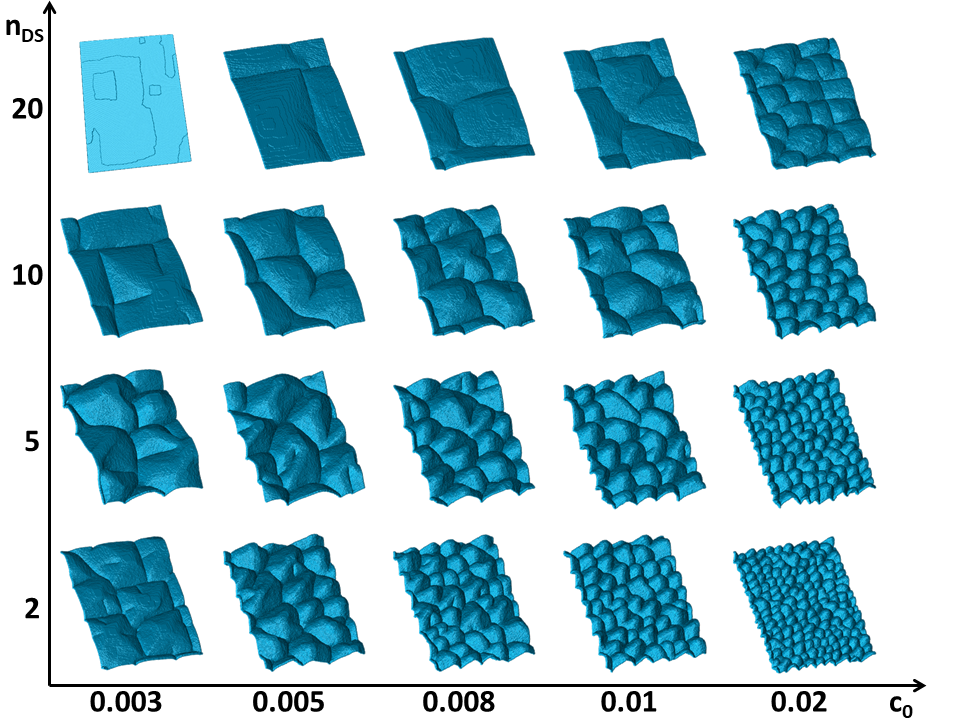}
\caption{Diagram of different surface patterns presented as a function of number of diffusional jumps $n_{DS}$ dependent on the initial particle concentration $c_0$ for flat surface, $E_V = 1.0 k_BT$  and $E_{ES} = 3.0 k_BT$. The structures are presented for the same number of layers for each $n_{DS}$ separately and equal to 282, 406, 267, 157 layers for $n_{DS}=$ 2, 5, 10 and 20 respectively. System size $300$ x $400$.}
\label{nds-vs-c0-L300}
\end{figure*}
\begin{figure*}[hbt]
\centering
a) \includegraphics[width=0.17\textwidth]{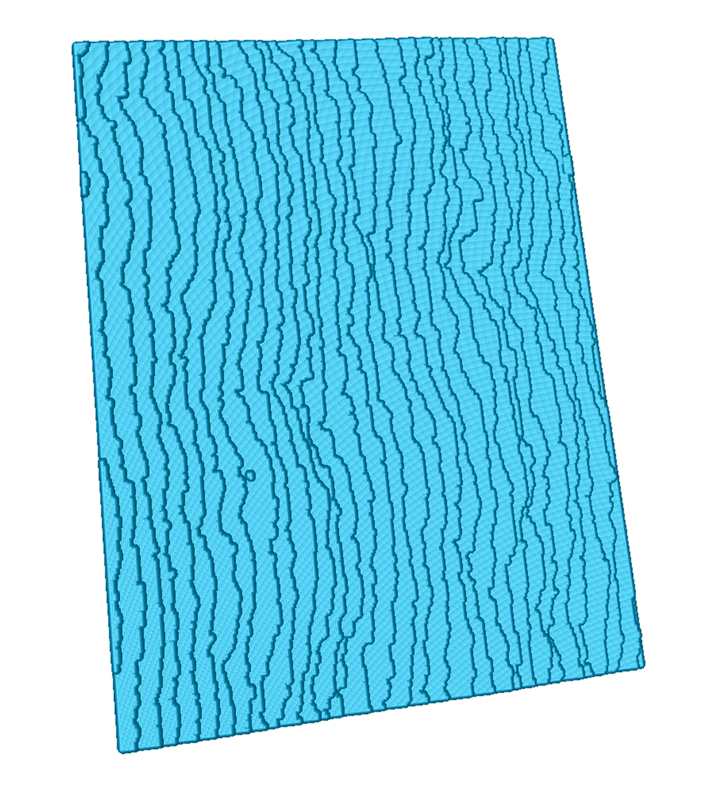}
b) \includegraphics[width=0.17\textwidth]{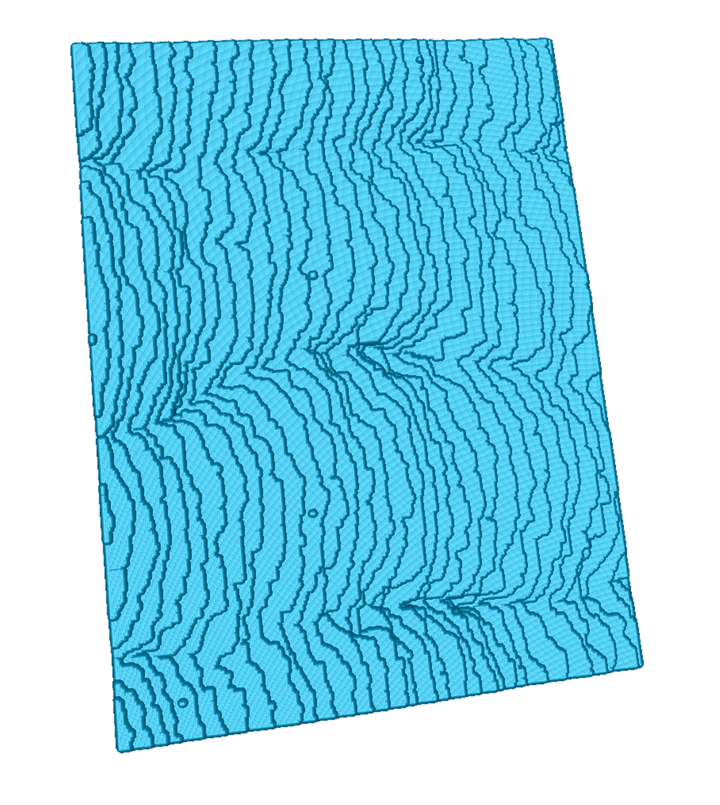}
c) \includegraphics[width=0.17\textwidth]{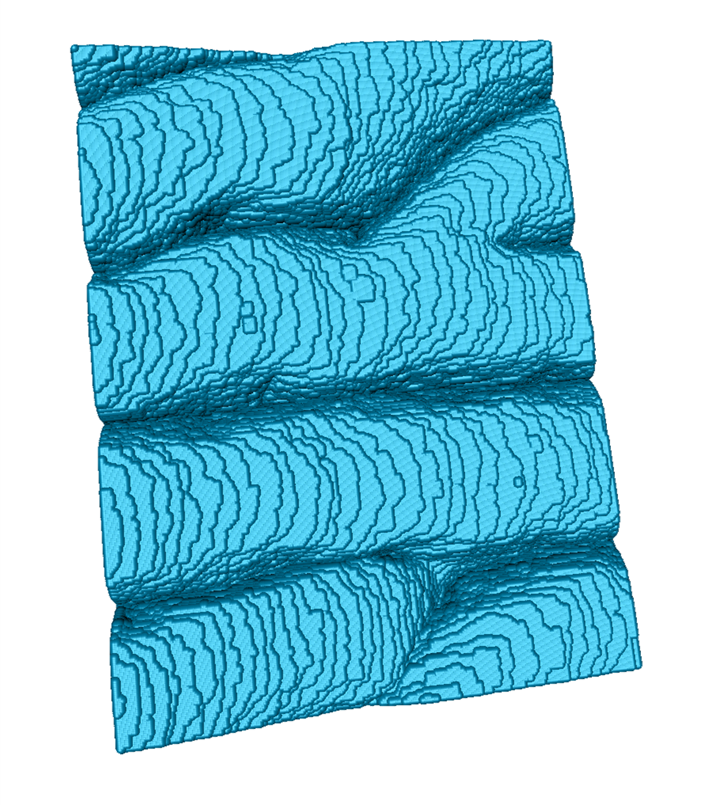}
d) \includegraphics[width=0.17\textwidth]{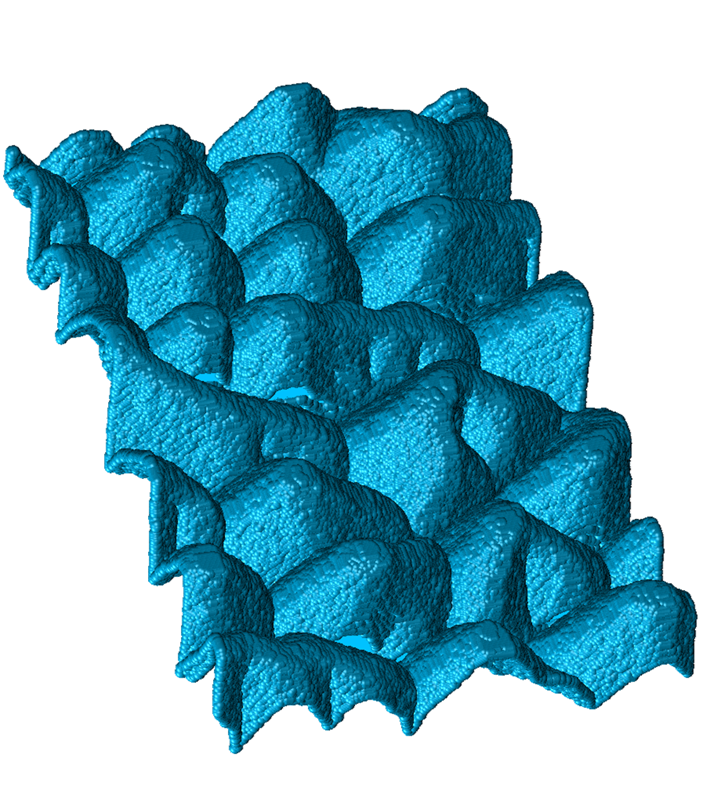}
e) \includegraphics[width=0.17\textwidth]{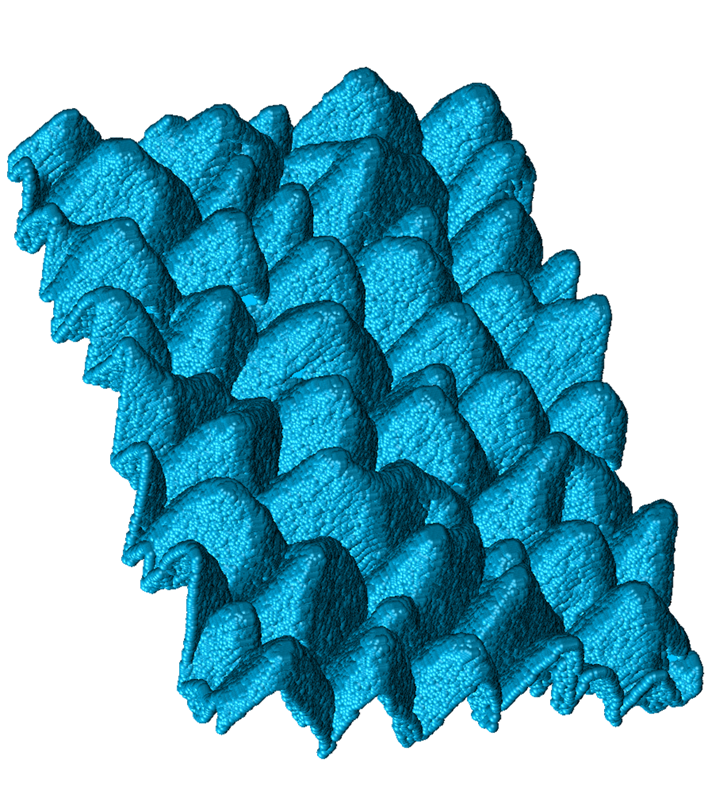}
\caption{Structures obtained for $n_{DS} = 5$, $c_0 = 0.01$, $l_0 = 10$, $E_V = 1.0 k_BT $ and a) $E_{ES} = 0.0$, b) $E_{ES} = 1.0 k_BT$, c) $E_{ES} = 2.0 k_BT$, d) $E_{ES} = 3.0 k_BT$, e) $E_{ES} = 4.0 k_BT$. Simulation time $10^6$. System size $300$ x $400$.}
\label{ES-dependence}
\end{figure*}
\begin{figure*}[hbt]
\centering
a) \includegraphics[width=0.16\textwidth]{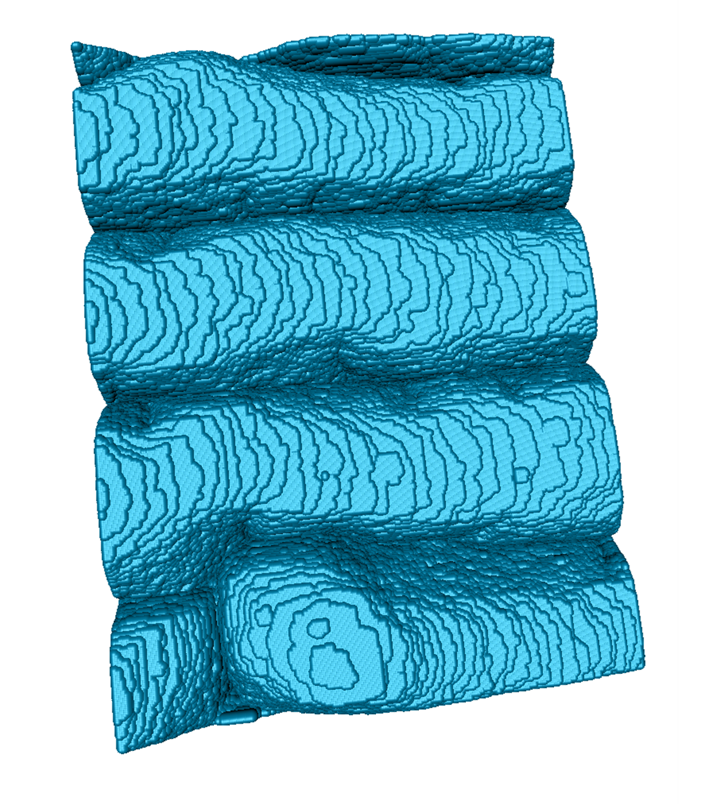}
b) \includegraphics[width=0.16\textwidth]{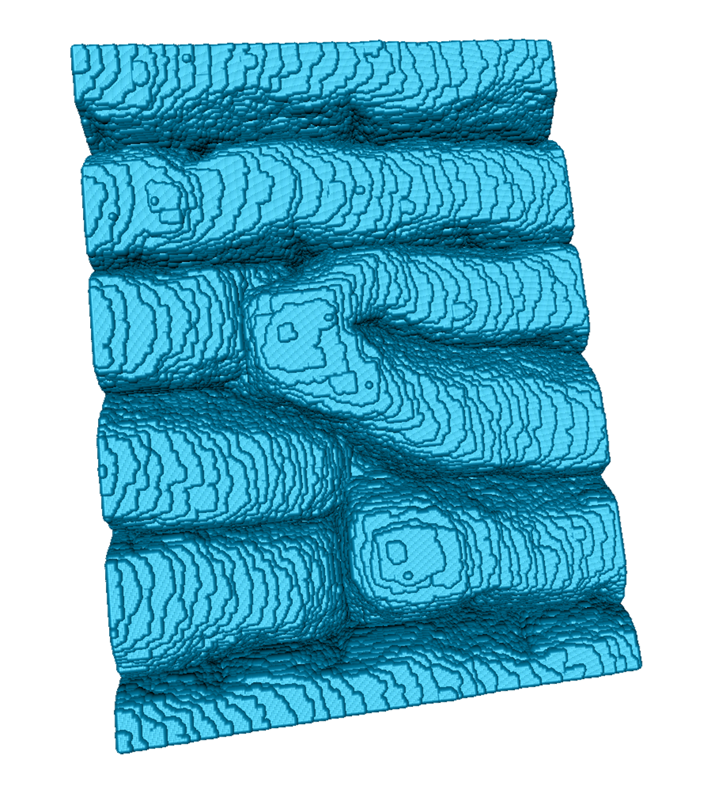}
c) \includegraphics[width=0.16\textwidth]{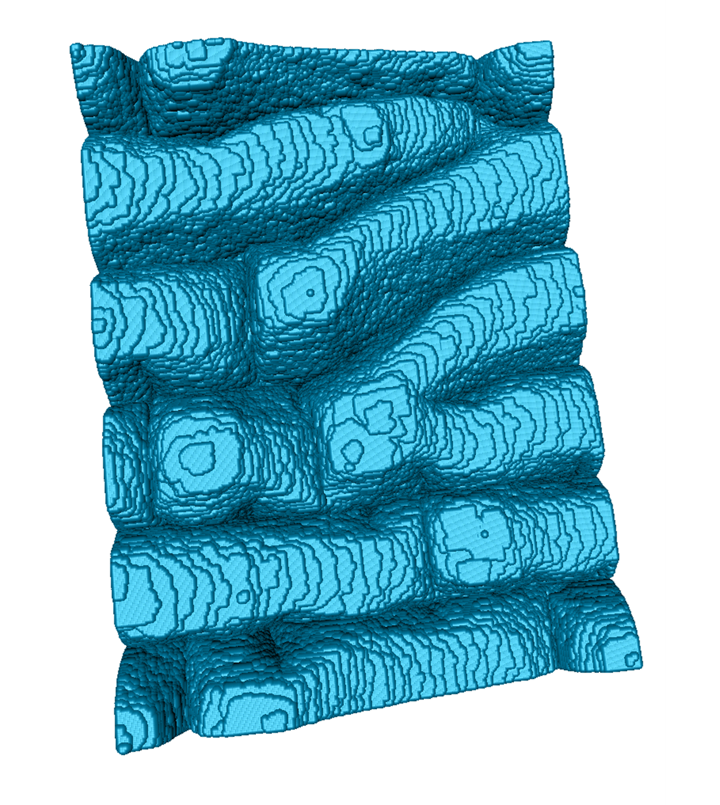}
d) \includegraphics[width=0.16\textwidth]{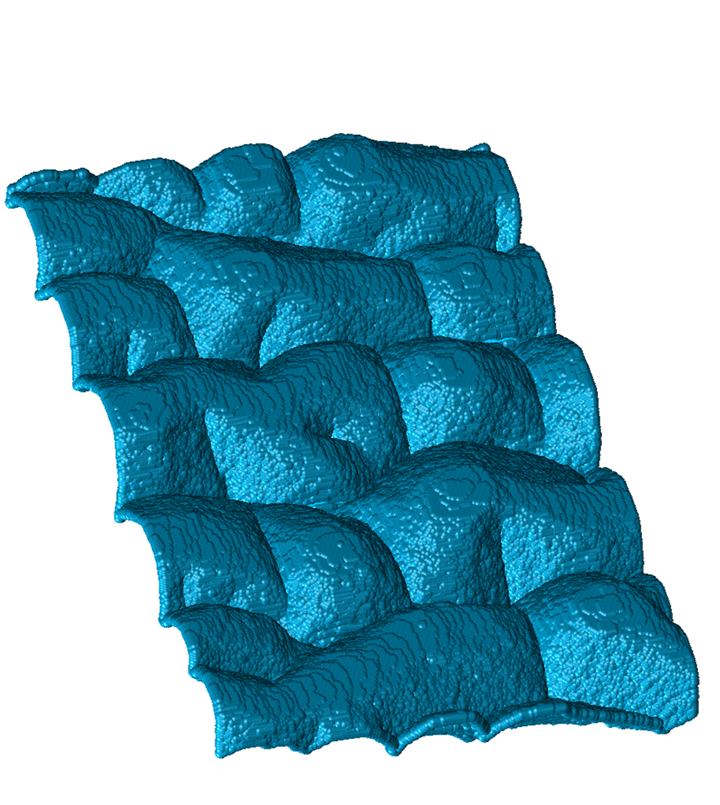}
e) \includegraphics[width=0.16\textwidth]{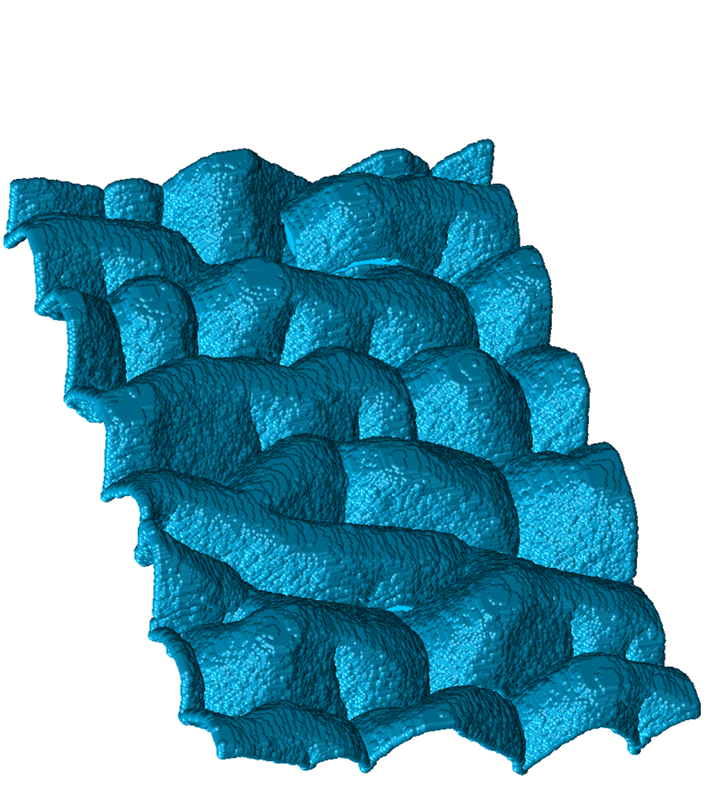}
\caption{Structures obtained for $n_{DS} = 5$, $c_0 = 0.01$, $l_0 = 10$, $E_V = 1.0 k_BT$, and a) $E_{ES} = 2.2 k_BT$, b) $E_{ES} = 2.4 k_BT$, c) $E_{ES} = 2.5 k_BT$, d) $E_{ES} = 2.6 k_BT$, e) $E_{ES} = 2.8 k_BT$. Simulation time $10^6$. System size $300 \times 400$.}
\label{ES-dense-dependence}
\end{figure*}

\begin{figure*}[hbt]
 \centering
a) \includegraphics[width=0.16\textwidth]{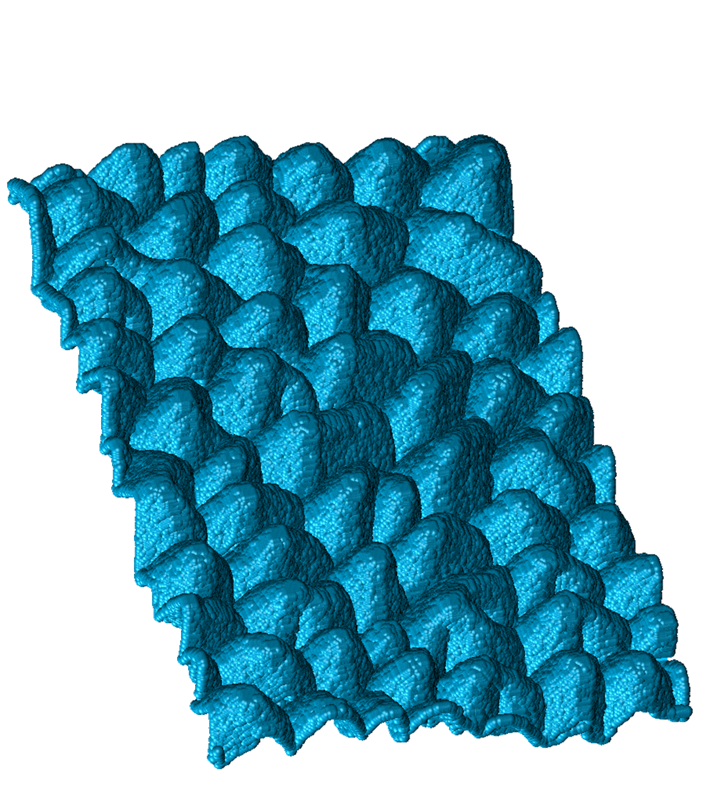}
b) \includegraphics[width=0.16\textwidth]{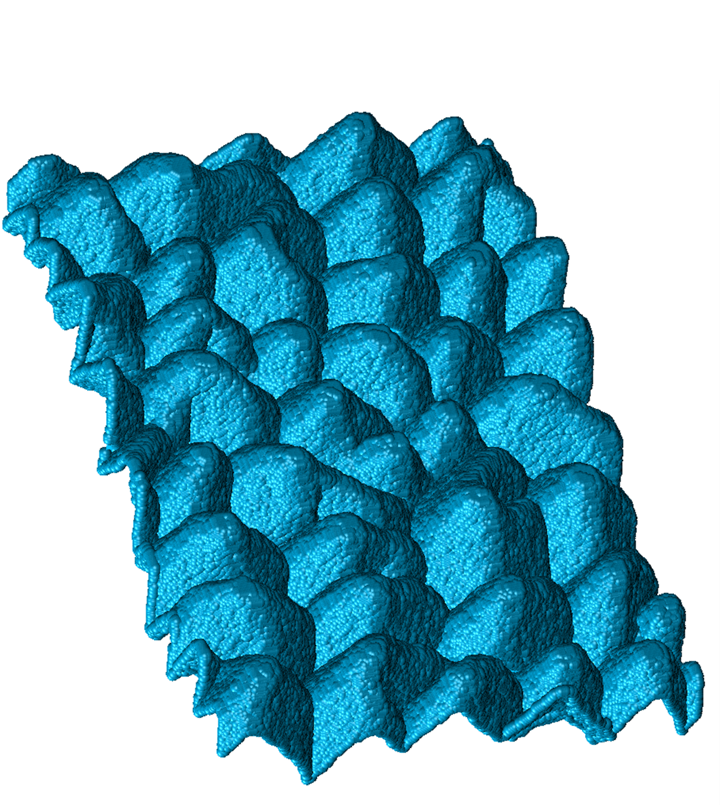}
c) \includegraphics[width=0.16\textwidth]{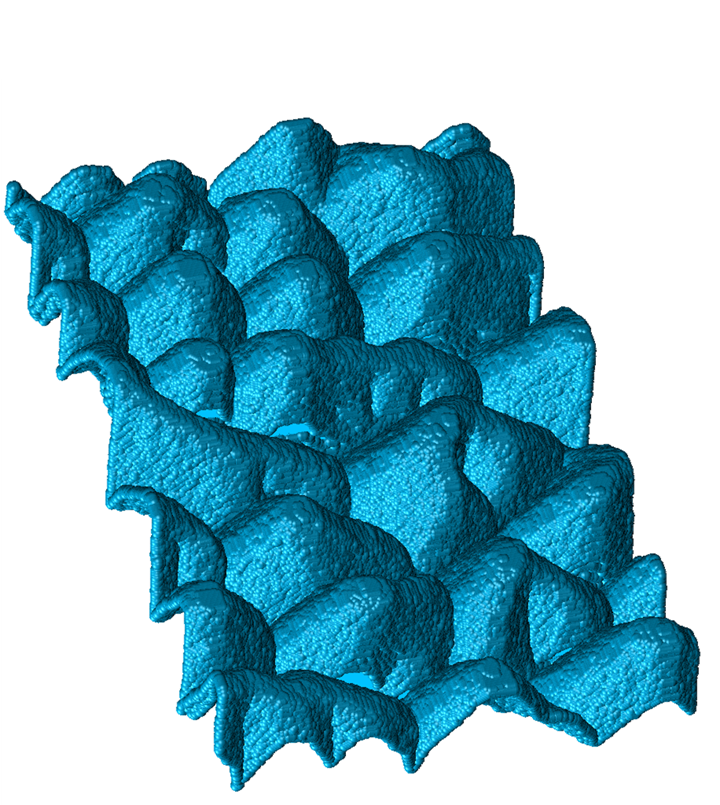}
d) \includegraphics[width=0.16\textwidth]{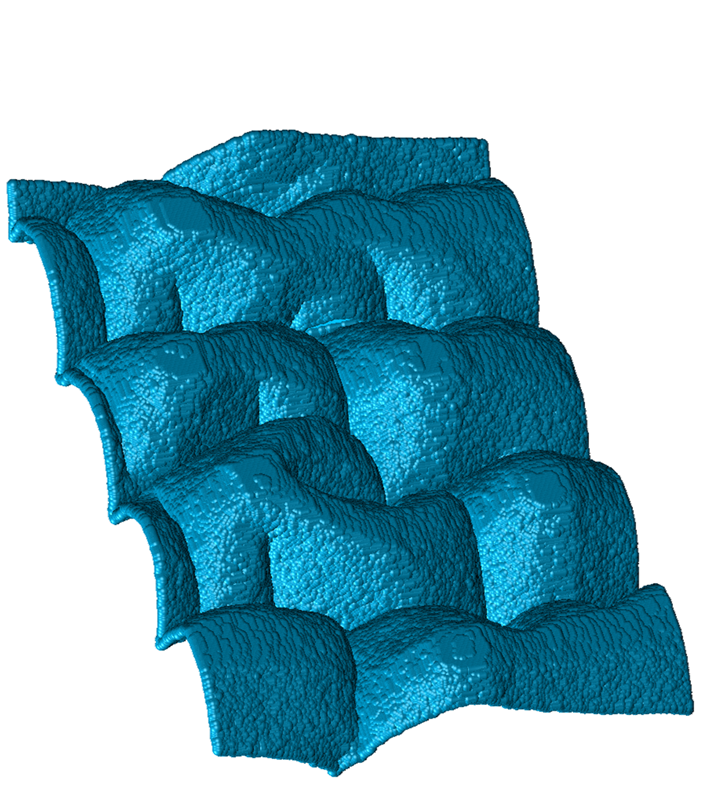}
e) \includegraphics[width=0.16\textwidth]{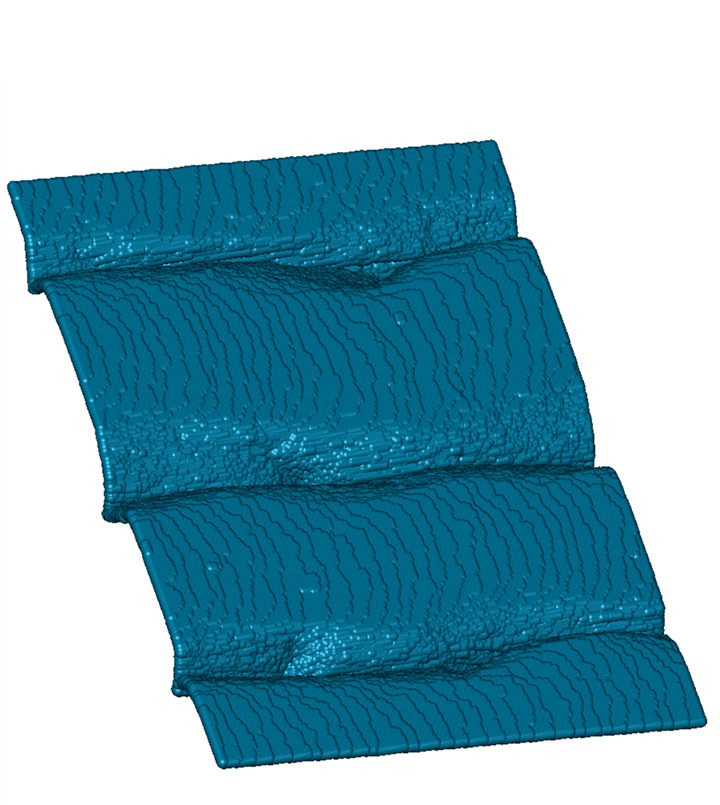}

\includegraphics[width=0.21\textwidth, height=0.17\textwidth]{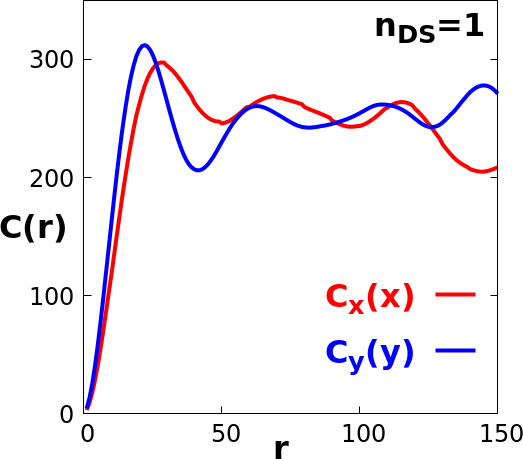}
\includegraphics[width=0.19\textwidth, height=0.17\textwidth]{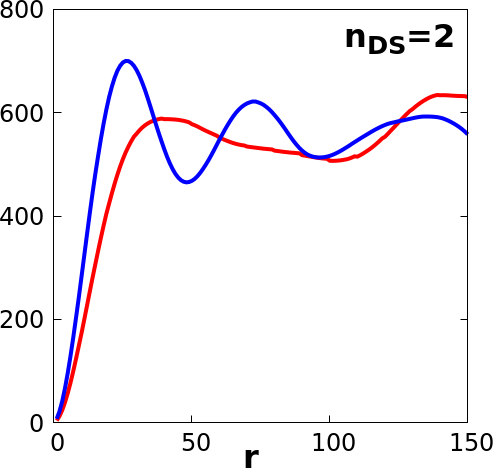}
\includegraphics[width=0.19\textwidth, height=0.17\textwidth]{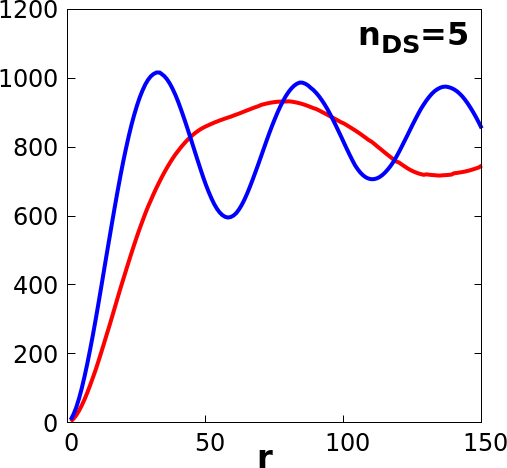}
\includegraphics[width=0.19\textwidth, height=0.17\textwidth]{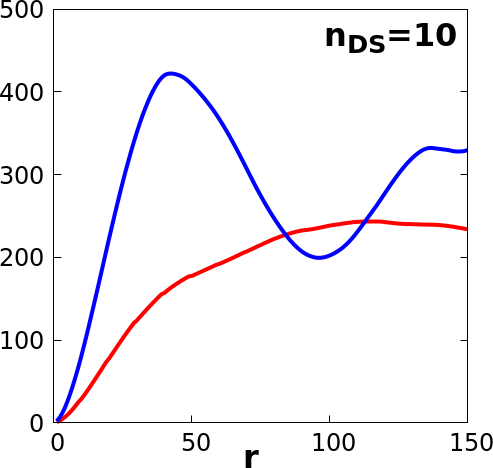}
\includegraphics[width=0.19\textwidth, height=0.17\textwidth]{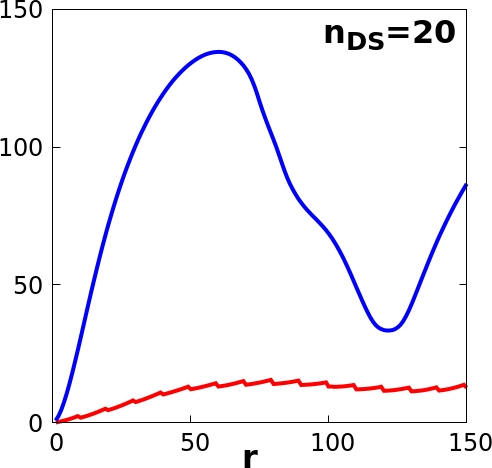}
\caption{Top panel: Structures obtained for $c_0 = 0.01$, $l_0 = 10$, $E_V = 1.0 k_BT$, $E_{ES} = 3.0 k_BT$ and a) $n_{DS} = 1$, b) $n_{DS} = 2$, c) $n_{DS} = 5$, d) $n_{DS} = 10$, e) $n_{DS} = 20$.
Bottom panel: Behavior of correlation functions $C_x$ and $C_y$, calculated along $x$-axis (across steps) and $y$-axis (along steps) respectively, for different number of diffusional jumps $n_{DS}$ corresponding to the morphological structures presented in the top panel. Simulation time $10^6$. System size $300$ x $400$.}
\label{nds-dependence}
\end{figure*}

 \subsection{Mound Morphologies on Flat Crystal Surfaces}
 At the flat surface, initially devoid of any steps, the growth process begins through island nucleation. In subsequent stages, nucleation occurs atop existing islands, and this behavior strongly depends on the height of the ES barrier, $E_{ES}$. A high barrier prevents adatoms from descending to lower layers. As a result, adatom density increases on the island tops, promoting upward growth.
The influence of the ES barrier height on the resulting surface morphology is illustrated in Fig.~\ref{ev-vs-es-L300}. All simulations were performed using the same values of $E_V=1k_BT$ and $c_0=0.01$, and the system was evolved for the same number of simulation steps, $t=10^6$. Results for different values of $n_{DS}$ are compared along the vertical axis.
A clear transition can be observed between $E_{ES}=2k_BT$ and $E_{ES}=3k_BT$, which appears consistently across all $n_{DS}$ values. Islands form in all cases, but at lower ES barriers, they remain relatively flat. These flatter islands tend to coalesce rather than grow vertically into mounds or pyramidal structures.
Closer examination reveals that $E_{ES}=1k_BT$ is a particularly distinct case. Although this does not imply a smooth surface potential, the resulting surface morphologies for all diffusion rates defined by $n_{DS}$ are limited to a height of one or, at most, two atomic layers. For $E_{ES}=2k_BT$, islands with greater vertical structure begin to emerge, though they are still relatively low and flat. Interestingly, new layers tend to form on top only when the uppermost terrace becomes nearly as large as the base layer.
In contrast, for all systems with $E_{ES} \geq 3k_BT$, islands grow vertically, developing into three-dimensional pyramidal structures with sharp tops. The morphology of these pyramids is strongly influenced by the surface diffusion rate: higher diffusion rates lead to the formation of larger pyramids with broader, more square-shaped bases, whereas lower diffusion rates result in smaller, steeper pyramids with rounded bases. A more detailed examination of the transition from $E_{ES}=2k_BT$ toward $E_{ES}=3k_BT$ is presented in Fig.~S1 in the Supplementary Material.

In Fig.~\ref{nds-vs-c0-L300}, we illustrate how the shape and size of the grown structures for $E_{ES}=3k_BT$ depend on the external flux-represented by $c_0$-and the parameter $n_{DS}$, which can be interpreted as an effective measure of diffusion rate, increasing with temperature. For each $n_{DS}$ separately, the results correspond to the same total number of deposited monolayers. Consequently, for lower adatom concentrations, the system was allowed to evolve for a proportionally longer time to reach the same growth stage.
It is evident that lower flux (smaller $c_0$) and higher temperatures (larger $n_{DS}$) favor the development of larger and more extended surface structures that tend to be relatively flat. Nonetheless, in all investigated cases, the surface consistently evolves into a regular mound-like morphology composed of individual units of varying sizes. Importantly, this results in a well-organized, albeit non-planar, surface structure rather than a rough or disordered one.

\subsection{Meander-to-Mound Morphological Transition on Vicinal Surfaces}
We observe that in all cases where growth takes place on an initially smooth surface, the most favorable mode of morphological evolution is the formation of mounds. In contrast, it is well established that the most effective strategy for achieving smooth, well-ordered crystal growth is to employ miscut (vicinal) surfaces. These surfaces, characterized by a regular array of atomic steps, serve as efficient sinks for adatoms and promote a stable, layer-by-layer growth mode. However, when the external particle flux is too high, even vicinal surfaces may lose their smooth morphology and develop mound-like features instead.
Simultaneously, it has been demonstrated that the presence of a potential well at the lower side of a step can trigger a strong meandering instability~\cite{Chabowska-PRB}. In general, the growth mode is sensitive to a number of factors, including kinetic barriers, terrace width, and diffusion rates.
In what follows, we show that by tuning only the height of the ES barrier, one can induce a smooth and continuous transformation of surface morphology - from regular meandered patterns to fully developed, faceted pyramidal structures. Furthermore, we investigate how these surface patterns are modulated by additional parameters such as the diffusion rate (quantified in our model by the number of diffusional jumps $n_{DS}$, which can also be interpreted as an effective temperature increase), terrace length, and the initial concentration of mobile adatoms.
Previous studies of meander formation~\cite{Chabowska-PRB} revealed that under certain conditions, localized mounds could begin to form at the crest of meandered structures. In this work, we focus specifically on that transitional regime and provide a detailed analysis of the emergence and evolution of such three-dimensional formations. To promote island growth, we employ a relatively high concentration of mobile adatoms ($c_0 = 0.01$) and explore the resulting surface morphologies over a range of ES barrier heights, $E_{ES}/k_BT \in [0.0, 8.0]$.

\begin{figure*}[hbt]
 \centering
a)\includegraphics[width=0.47\textwidth]{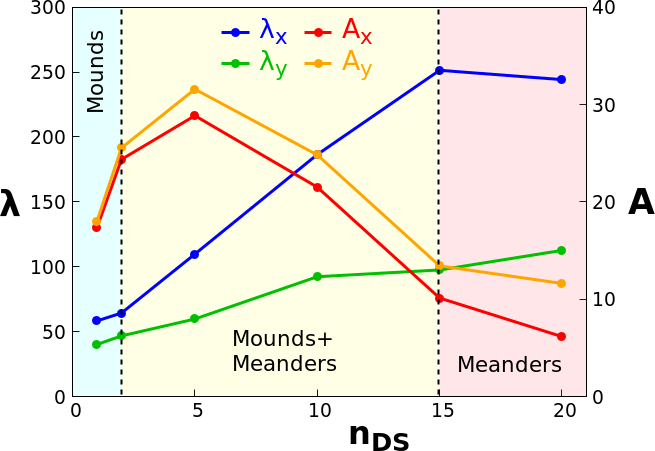}
\hspace{0.2 cm}
b)\includegraphics[width=0.47\textwidth]{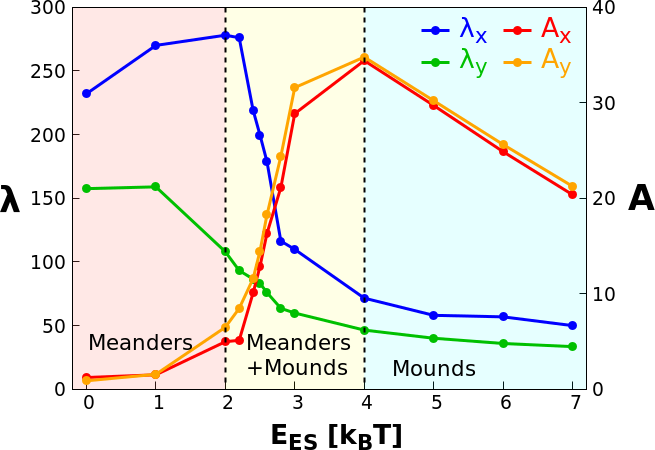}
\caption{Influence of a) diffusion rate $n_{DS}$ and b) Ehrlich-Schwoebel barrier height $E_{ES}$ on the main morphological characteristics $\lambda$ and $A$, calculated from correlation functions along $x$- and $y$-axis separately, distinguishing the structures grown on vicinal surface. Other system parameters are: $c_0 = 0.01, l_0=10, E_V = 1 k_BT, E_{ES} = 3 k_BT$ in (a), $n_{DS}=5$ in (b), simulation time $10^6$, and system size $300$ x $400$. Regions corresponding to different morphological structures are marked in different colors.}
\label{fig_new-lambda-amplitude-ES-nds-influence}
\end{figure*}

Figure~\ref{ES-dependence} presents a representative set of patterns obtained under fixed conditions, showing the progressive morphological evolution as the ES barrier increases. As seen in Fig.~\ref{ES-dependence}, in the absence of an ES barrier, the final surface exhibits a regular arrangement of steps, typical of stable vicinal growth. Introducing a small barrier leads to the onset of step meandering, which becomes increasingly pronounced as the barrier height increases.
At intermediate barrier values, we observe a qualitative change in surface morphology: the regular meanders give way to the formation of 3D mounds. Further increasing $E_{ES}$ results in enhanced vertical growth and sharper, more faceted pyramidal structures. These observations underline the crucial role of the ES barrier in controlling the transition from two-dimensional instabilities to fully developed three-dimensional surface features.
The change in the shape of the surface structures is observed to occur within the range of $E_{ES}$ between $2$ and $3$. To better understand this transition, we performed a more detailed analysis within this interval of barrier heights. The corresponding results are presented in Figure~\ref{ES-dense-dependence}.
As evident from the figure, the transition from meandered patterns to pyramidal structures occurs in a continuous and gradual manner. As the height of the ES barrier increases, small islands begin to form on the crests of the meanders. These islands act as nucleation sites for the vertical growth of three-dimensional structures. Initially, the surface exhibits a hybrid morphology-intermediate between meanders and pyramids-as shown in Figures~\ref{ES-dense-dependence}c-e. With further increase of $E_{ES}$, the structures become more distinctly three-dimensional, as seen in Figure~\ref{ES-dependence}e, marking the full development of faceted pyramidal features.

The transformations discussed above were obtained for a relatively low diffusion rate, with $n_{DS} = 5$. In Fig.~\ref{nds-dependence}, we examine how increasing the number of diffusional jumps, $n_{DS}$, affects the resulting surface morphologies. Specifically, we investigated the evolution of 3D structures as a function of $n_{DS}$ in the range from $1$ to $20$.
As shown in the top panel of Fig.~\ref{nds-dependence}, increasing the diffusion rate has a significant impact on the surface morphology. At low values of $n_{DS}$, the system forms densely packed, steep pyramidal structures. As $n_{DS}$ increases, these structures begin to elongate and lose their vertical symmetry. Eventually, for sufficiently high diffusion rates, the pattern evolves back into a meander-like configuration. This recovery of meanders at high diffusivity highlights the delicate balance between the ES barrier and adatom mobility in determining the final surface morphology.

\subsection{Characterization of pattern formation by using the height-height correlation function}
The height-height correlation function $C(r)$ is used as a powerful tool for characterizing surface patterns. It measures the average squared height difference between two surface points $i$ and $j$, separated by a distance $r = r_i - r_j$, and is defined as:
\begin{equation}
\begin{split}
C(r) = C(r_i - r_j) = \langle [h(r_i) - h(r_j)]^2 \rangle = \\
= \langle [h(x_i,y_i) - h(x_j,y_j)]^2 \rangle .
\label{Eq:corr_function}
\end{split}
\end{equation}

In order to characterize the anisotropic evolution of the surface morphology, it is convenient to evaluate the correlation function along the $x$ and $y$ directions separately using the following expressions:
\begin{equation}
C_x(x) = \frac{1}{(L_x-x)L_y} \sum_{k = 1}^{L_x-x} \sum_{l = 1}^{L_y} [h(k+x,l) - h(k,l)]^2 ,
\label{Eq:corr_function_x}
\end{equation}

\begin{equation}
C_y(y) = \frac{1}{L_x(L_y-y)} \sum_{k = 1}^{L_x} \sum_{l = 1}^{L_y-y} [h(k,l+y) - h(k,l)]^2 ,
\label{Eq:corr_function_y}
\end{equation}
where $C_x$ and $C_y$ denote the correlation functions in the directions across and along the steps, respectively; $L_x = 300$ and $L_y = 400$ are the system sizes; $x$ and $y$ are distances varying from $1$ up to half the system size in the corresponding direction. Each correlation function is averaged over the orthogonal direction, ensuring full sampling of the surface.

Two characteristic lengths of the surface morphology can be extracted from the correlation function: the wavelength $\lambda$ and the amplitude $A$. These quantities describe the lateral periodicity and vertical modulation of the surface, respectively. For meandered structures, the most relevant parameter is the wavelength $\lambda$, whereas mounded structures are characterized by both their lateral size (given by the correlation length) and their amplitude $A$. The wavelength $\lambda$ is determined as the position of the first minimum of $C(r)$, while the amplitude $A$ corresponds to its first maximum. Both quantities are calculated separately along the $x$ and $y$ directions. In addition, the results are averaged over ten independent realizations.

The behavior of correlation functions $C_x$ and $C_y$, computed along the $x$ and $y$ axes, is presented in the bottom panel of Fig.~\ref{nds-dependence} for structures obtained at different numbers of diffusional jumps $n_{DS}$, varying from $1$ to $20$. These results correspond to the surface morphologies shown in the top panel of the same figure. As $n_{DS}$ increases, the surface morphology evolves from compact mounds, through an intermediate state, towards well-developed meanders. This transformation is clearly reflected in the correlation functions: initially isotropic mounds exhibit nearly identical $C_x$ and $C_y$, while intermediate states display similar amplitudes but different correlation lengths. For high $n_{DS}$ values, the functions separate into two distinct behaviors - $C_y$ with a pronounced amplitude and wavelength, and $C_x$ nearly flat - characteristic of a regular step meandered pattern. Similar shape of the height-height correlation function was shown in Ref.~\cite{Evans} for mounded structures. Additionally, in Fig.~S2 in the Supplementary Material we present a comparison of the correlation functions and the corresponding structures grown at different diffusion rates $n_{DS}$ on vicinal and flat surfaces, respectively.

The dependence of the main morphological characteristics $\lambda$ and $A$ on the two key parameters - the Ehrlich-Schwoebel barrier height $E_{ES}$ and the diffusion rate $n_{DS}$ - is summarized in Fig.~\ref{fig_new-lambda-amplitude-ES-nds-influence}. Three regimes can be identified from these plots: meandered, intermediate, and mounded.
In contrast, the analogous dependence obtained for a flat surface (presented in Fig.~S3) reveals only the mounded regime. As can be seen, the mounds are characterized with almost the same wavelengths and amplitudes in both directions, typically for isotropic structures. In the intermediate regime,  when a mixture of mounds and meanders emerges, both morphological characteristics $\lambda$ and $A$ change drastically and synchronically in both directions. Whereas the regular meanders can be clearly identified in the significant difference between the wavelengths $\lambda_x$ and $\lambda_y$ along both directions and the small amplitudes, underlining the anisotropic character of the structure.

In addition, the time evolution of the correlation functions for regular mounded, regular meandered and mixed structure is provided in the Supplementary Material (see Fig.~S4). Moreover, the time-scaling behavior of the main characteristic lengths along both directions - the wavelengths $\lambda_x$ and $\lambda_y$, and the amplitudes $A_x$ and $A_y$, is presented for different diffusion rate $n_{DS}$ in Fig.~S5 in the Supplementary Material. Their evolution can be described by the following scaling relations:
\begin{eqnarray}
\lambda &\sim& t^{1/z}, \\
A &\sim& t^{\beta} ,
\end{eqnarray}
with scaling exponents $1/z$ and $\beta$ listed in Tables~\ref{tab:tab-1} and \ref{tab:tab-2}, respectively, for structures grown at different diffusion rates $n_{DS}$ on both vicinal and flat surfaces. Note that for meandered structures grown on vicinal surface the correlation function across the step direction exhibits very small amplitude and an ill-defined wavelength (see Fig.~\ref{nds-dependence}e, bottom panel). However, $\lambda_y$ and $A_y$ measured along step direction allow for distinguishing the different growth regimes, highlighting the contrasting dynamical evolution of meanders and mounds. According to Ref.~\cite{Golubovic}, in the case of a square lattice, the coarsening exponent - also referred as the dynamic exponent 1/z - should be equal to $0.25$.  Indeed, we obtained this value for a flat surface. Rost et al. claim in their paper \cite{Rost} that the same value should apply to flat and vicinal surfaces. Based on our results, we observe that this only occurs  in mounded and mixed regions. For high values of $n_{DS}$, for which the meanders are observed on vicinal surface, the exponents are obviously different. This is correlated with elongation of the structure in one direction, which causes more distinguishable directions for meanders. Traces of the influence of direction are also visible in the other two regions, particularly in the mixed region for the vicinal surface. The analysis of the growth exponent $\beta$, which is correlated with the roughening exponent found in the literature, reveals different growth behavior for vicinal surfaces. Mounded structures grow similarly on vicinal and flat surfaces. However, when the growth regime changes to mixed or meandered, the growth exponents also change.

\begin{table}
\centering
        \begin{tabular}{|c|c|c||c|}
            \hline
              &  \multicolumn{3}{c|}{${1/z}$}   \\
            \cline{2-4}
            \multirow[c]{3}{*}{$n_{DS}$} & \multicolumn{2}{c||}{vicinal surface} & flat surface\\
            \cline{2-4}
              & $x$ & $y$ &  $x =y$ \\
            \hline
            \rowcolor{cyanish} %
             1 & 0.33 & 0.2 & 0.2\\
            \hline
            \rowcolor{cyanish} %
             2 & 0.33 & 0.2 & 0.2 \\
            \rowcolor{yellowish} %
            \hline
             5 & 0.5 & 0.25 & \cellcolor{cyanish} 0.25 \\
            \rowcolor{yellowish} %
            \hline
            10 & 0.5 & 0.25 & \cellcolor{cyanish} 0.25 \\
            \rowcolor{pinkish} %
            \hline
            15 & 0.75 & 0.4 & \cellcolor{cyanish} 0.25 \\
            \rowcolor{pinkish} %
            \hline
            20  & 0.75 & 0.4 & \cellcolor{cyanish} 0.25 \\
            \hline
        \end{tabular}
        \caption{The dynamic exponent $1/z$ calculated separately across (x) and along (y) the steps for structures grown at different diffusion rates $n_{DS}$ on vicinal and flat surfaces. The margin of error is 0.05.}
        \label{tab:tab-1}
\end{table}
\begin{table}
      \centering
        \begin{tabular}{|c|c|c||c|}
            \hline
              &  \multicolumn{3}{c|}{$\beta$}   \\
            \cline{2-4}
            \multirow[c]{3}{*}{$n_{DS}$}  &  \multicolumn{2}{c||}{vicinal surface} & flat surface\\
            \cline{2-4}
              &  $x$ & $y$ & $x = y$\\
            \hline
            \rowcolor{cyanish} %
             1 & 1.8 & 1.8 & \cellcolor{cyanish} 1.8\\
            \hline
            \rowcolor{cyanish} %
             2 & 1.8 & 1.8 & \cellcolor{cyanish} 1.9\\
            \hline
            \rowcolor{yellowish} %
             5 & 1.25 & 1.25 & \cellcolor{cyanish} 2\\
            \hline
            \rowcolor{yellowish} %
            10 & 1.25 & 1.25 & \cellcolor{cyanish} 2\\
            \hline
            \rowcolor{pinkish} %
            15 & 1.1& 1.1 & \cellcolor{cyanish} 2\\
            \hline
            \rowcolor{pinkish} %
            20  & 0.75 & 1.1 & \cellcolor{cyanish} 2 \\
            \hline
        \end{tabular}
        \caption{The growth exponent $\beta$ calculated separately across (x) and along (y) the steps for structures grown at different diffusion rates $n_{DS}$ on vicinal and flat surfaces. The margin of error is 0.05.}
        \label{tab:tab-2}
\end{table}

\begin{figure*}[hbt]
 \centering
\includegraphics[width=0.8\textwidth]{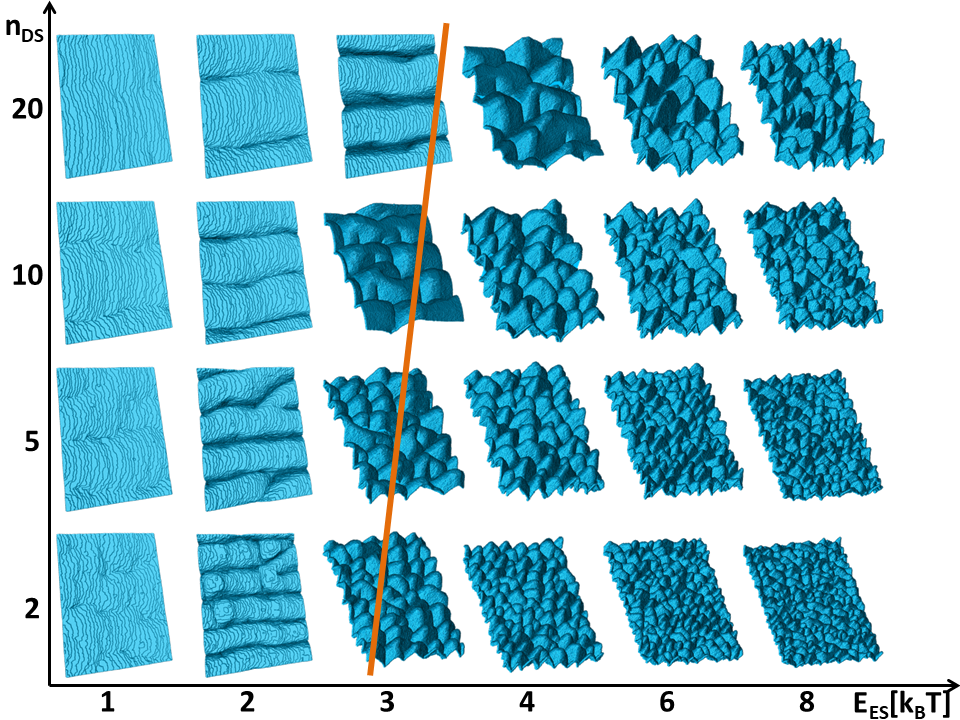}
\caption{Diagram of different surface patterns presented as a function of number of diffusional jumps $n_{DS}$ dependent on the height of ES barrier $E_{ES}$ for $c_0 = 0.01$, $l_0 = 10$, and $E_V = 1.0 k_BT$. Simulation time $10^6$. System size $300$ x $400$. The presented orange line is obtained from Eq.~\ref{boundary_line}.}
\label{nds-vs-ES}
\end{figure*}

\begin{figure*}[hbt]
 \centering
\includegraphics[width=0.8\textwidth]{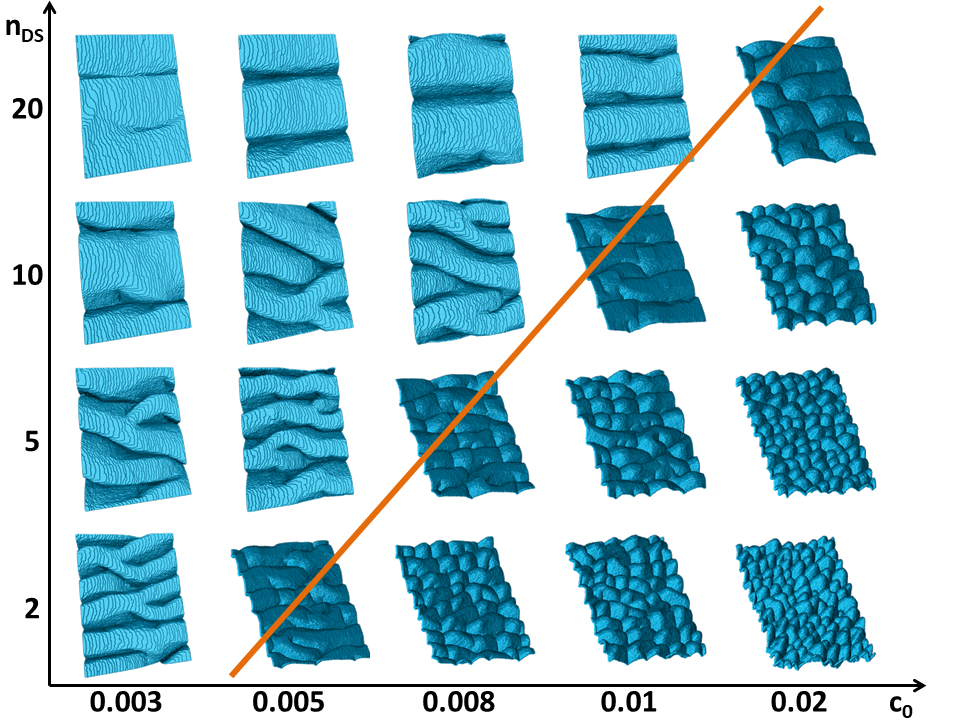}
\caption{Diagram of different surface patterns presented as a function of number of diffusional jumps $n_{DS}$ dependent on the initial particle concentration $c_0$ for $l_0 = 10$, $E_V = 1.0 k_BT$ and $E_{ES} = 3.0 k_BT$. The structures are presented for the same number of layers for each $n_{DS}$ separately and equal to 293, 357, 372, 449 layers for $n_{DS}=$ 2, 5, 10 and 20 respectively. System size $300$ x $400$. The presented orange line is obtained from Eq.~\ref{boundary_line}. }
\label{nds-vs-c0-L10}
\end{figure*}

\begin{figure*}[hbt]
 \centering
\includegraphics[width=0.8\textwidth]{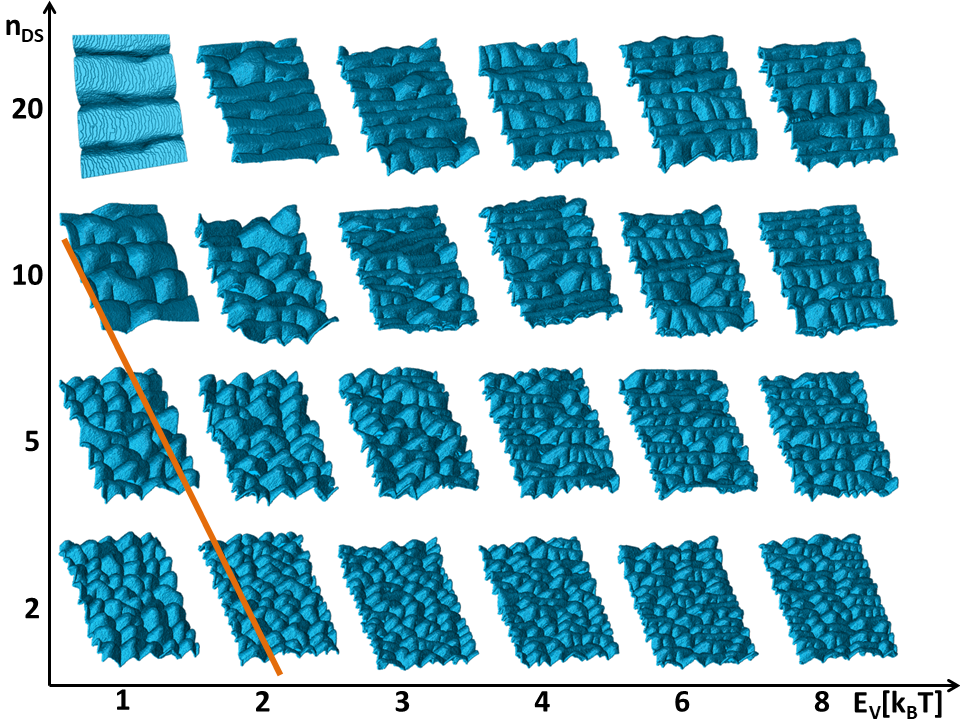}
\caption{Diagram of different surface patterns presented as a function of number of diffusional jumps $n_{DS}$ dependent on the depth of the potential well $E_V$ for $l_0 = 10$, $c_0 = 0.01$ and $E_{ES} = 3.0 k_BT$. Simulation time $10^6$. System size $300$ x $400$. The presented orange line is obtained from Eq.~\ref{boundary_line}.}
\label{nds-vs-EV}
\end{figure*}

\subsection{Morphology  diagrams}

 A key observation can be summarized as follows: the meandered surface patterns evolves into a mounded landscape as the height of the ES barrier $E_{ES}$ increases. Remarkably, this trend can be reversed by enhancing the surface diffusion. In other words, a system that develops mounded structures at moderate diffusion rates may revert to a meandered morphology once diffusion becomes sufficiently fast.

This interplay is captured in Fig.~\ref{nds-vs-ES}, which maps the observed surface patterns as a function of the number of diffusional jumps $n_{DS}$ and the ES barrier height $E_{ES}$. As the diagram shows, the critical ES barrier required for the meander-to-mound transition shifts to higher values with increasing $n_{DS}$. For small $E_{ES}$, meandered structures always appear, irrespective of the diffusion rate. The transition line separating the two regimes can be approximately expressed as:
\begin{equation}
\sqrt{n_{DS}} \exp(- 2 \beta E_{ES}) = \mathrm{const}.
\end{equation}
This scaling relation highlights that the destabilizing effect of the ES barrier can be counterbalanced by enhanced adatom mobility. At large $n_{DS}$, a particle can perform many surface jumps before incorporation into the crystal, effectively overcoming even a sizable ES barrier. In this way, higher diffusion not only increases lateral transport but also enhances step permeability, enabling adatoms to cross step edges instead of being trapped. As a result, the system recovers meander-like morphologies instead of mounded ones.

Another key parameter that significantly influences surface nucleation-and therefore plays a central role in pattern formation-is the adparticle density $c_0$. This parameter can be interpreted as the flux of particles arriving at the surface. In Fig.~\ref{nds-vs-c0-L10}, we present how the surface morphology evolves as a function of both the number of diffusional jumps $n_{DS}$ and the initial adparticle concentration $c_0$. The simulation time was normalized such that the same number of monolayers was deposited in each case.
The diagram reveals that higher values of $c_0$ tend to promote mound formation, while increasing $n_{DS}$ leads to smoother surfaces and supports the emergence of meandered structures. The transition between these two growth regimes-meanders and mounds-can be approximately described by the relation:
\begin{equation}
\sqrt{n_{DS}} c_0^{-1} = \mathrm{const},
\end{equation}
indicating that $c_0$ acts in a manner analogous to the ES barrier. A higher particle flux (i.e., higher $c_0$) reduces the effective mobility of adatoms, favoring mound formation in a way similar to an increased ES barrier.
Interestingly, surface patterns located near the transition line (just below the meander-to-mound boundary) exhibit a characteristic and reproducible morphology. These structures typically consist of rectangular mounds, elongated in the direction perpendicular to the steps, indicating a well-defined intermediate regime between meanders and fully developed pyramids.

In all of the previously discussed examples, we kept the depth of the potential energy well fixed at $E_V =1~k_BT$. Generally, the primary control parameter governing the transition between meandered and mounded surface morphologies is the height of the ES barrier, $E_{ES}$, rather than the depth of the potential well, $E_V$. However, $E_V$ significantly influences the wavelength of the meanders, as analyzed in detail in Ref.\cite{Chabowska-PRB}. In Fig.\ref{nds-vs-EV}, we illustrate how variations in the potential well depth affect the shape of mounds and alter the characteristics of the intermediate regime between meanders and mounds.
It can be observed that for high diffusion rates (i.e., large $n_{DS}$), the surface remains in a meandered state even at large values of $E_V$. Nevertheless, increasing $E_V$ leads to a noticeable decrease in the meander wavelength and the appearance of secondary wave-like modulations along the sides of the primary finger-like structures. Interestingly, similar meander morphologies are found even at lower diffusion rates, provided the potential well is sufficiently deep. Regular mound-like structures reappear only when the diffusion rate is very low and/or the potential well is shallow.

In order to understand the relations between the parameters described above, let us analyze the condition under which the transition from meanders to a mound structure occurs. The structure is assumed to remain meandering if the time needed to nucleate a second-layer island is longer than the time it takes for a step from one side and an island from the other side to move a distance of $l_0/2$ along a terrace and merge.
The time required for an island to grow can be approximated by the inverse of the nucleation probability, which is equal to  $\rho_{is}^4$, where $\rho_{is}$ is  adparticle  density on  top  of  the  island . In the presence of the ES barrier, we have $ \rho_{is} \sim c_0\exp(\beta E_{ES}) $ , because particles are continuously added  and are confined between barriers until they nucleate. On the other hand, the step moves forward with a velocity $ \rho_V^2$, which is the same as the island growth velocity, and depends on the density at the bottom of the step, $\rho_V$.

Combining these factors, we obtain:
\begin{equation}
    \frac{1}{c_0^4 \exp(4\beta E_{ES})} \geq \frac{l_0}{4 \rho_V^2},
\end{equation}

By applying the approximate  relation $\rho_V \sim \sqrt{n_{DS}} c_0 \exp(\beta E_V)$   , the expression simplifies to:
\begin{equation}
    \frac{2 \sqrt{n_{DS}} \exp(\beta E_V)}{c_0 \exp(2\beta E_{ES})\sqrt{l_0}} \geq 1.
    \label{boundary_line}
\end{equation}

The condition above defines the dividing line between meandered and mounded structures, accounting for both previously observed relations. Their form aligns well with the dependencies proposed above. We present the boundaries defined by Eq. \ref{boundary_line} in Figures \ref{nds-vs-ES}, \ref{nds-vs-c0-L10}, and \ref{nds-vs-EV}. In the first two diagrams, these lines fit the boundary between the meandered and mounded regions very well.

However, the third relation ($n_{DS}$ vs. $E_V$) differs from the others; it fails to properly separate the discussed regions of surface patterns, particularly at higher values of the bottom step potential, $E_V$. This suggests that for these parameters, Eq. \ref{boundary_line} does not fully capture the mechanism of three-dimensional growth. This discrepancy may be caused by the rapid meandering process in this region of the diagram, which leads to the formation of long, thin, finger-like structures separated by deep valleys. Such a process deviates significantly from the slow, monotonous forward step movement assumed previously.

At larger values of $E_V$, nucleation occurring on top of and between these "fingers"-rather than on flat terraces-presumably becomes the dominant process governing the final surface pattern. This indicates that the relationship between $E_V$ and the other parameters is increasingly complex. Conversely, at low $E_V$ values, $n_{DS}$ maintains a linear dependence on $l_0$. This linear relationship remains consistent with the observations made throughout our investigations.

\section{Conclusions}

In this work, we have systematically investigated surface pattern formation during epitaxial growth, focusing on the transition between meandered and mounded morphologies and vice versa. Using VicCA simulations, we examined how surface evolution depends on four key parameters: the ES barrier height $E_{ES}$, the number of diffusional jumps $n_{DS}$ representing surface mobility, the adatom concentration $c_0$ linked to deposition flux, and the potential well depth $E_V$ reflecting the strength of interatomic interactions. The transition from mounds to meanders is reflected in the anisotropy of the height-height correlation function used for quantitatively describing the surface morphologies.

We showed that increasing $E_{ES}$ suppresses adatom descent, promoting vertical growth and driving the transition from smooth, meandered structures to fully developed pyramidal mounds. This transition occurs gradually and is accompanied by the appearance of intermediate morphologies, such as elongated mounds oriented perpendicular to the steps. The competition between surface diffusion and the ES barrier is decisive: increasing $n_{DS}$ restores meandered structures even at high $E_{ES}$, demonstrating that enhanced surface mobility counteracts kinetic roughening.

A similar trend is observed with varying adatom flux. Increasing $c_0$ favors mound formation, while higher $n_{DS}$ has the opposite effect. Although $E_V$ does not directly control the meander-to-mound transition, it strongly influences the wavelength and modulation of meander patterns, especially at high diffusion rates.

Altogether, these results reveal that a wide spectrum of surface morphologies-from step patterns and meanders to sharp three-dimensional pyramids-can be systematically tuned by adjusting basic growth parameters. In particular, the existence of an intermediate morphological regime offers a potential pathway for nanoscale surface design.

The scaling laws and transitions identified here provide experimentally testable predictions for molecular beam epitaxy (MBE) and related growth techniques. In practice, $E_{ES}$ can be tuned through material choice or surfactant-mediated growth, while $n_{DS}$ and $c_0$ can be controlled by adjusting temperature and deposition flux, respectively. This establishes a direct link between our simulations and realistic experimental conditions, enabling exploration of the predicted morphological phase diagram.

Future work may extend this study by incorporating additional physical effects such as step permeability, anisotropic diffusion, or strain. In particular, the coupling between elastic interactions and step meandering remains an open problem, with implications for thin-film stress relaxation and pattern selection. Experimental validation, for instance through real-time surface imaging, would provide valuable confirmation of our predictions and could ultimately guide the rational design of nanoscale surface architectures via kinetic control.

\section*{Acknowledgments}
The authors express their gratitude to The Polish National Center for Research and Development (grant no. EIG CONCERT-JAPAN/9/56/AtLv-AlGaN/2023), The Bulgarian National Science Fund (grant No. KP-06-DO02/2/18.05.2023), the Polish Academy of Sciences and the Bulgarian Academy of Sciences (grant No. IC-PL/07/2024-2025) for providing financial support. They also extend their thanks to Vesselin Tonchev from the Faculty of Physics at Sofia University and Yoshihiro Kangawa from the Research Institute for Applied Mechanics at Kyushu University for their valuable discussions. Most of the calculations were done on HPC facility Nestum (BG161PO003-1.2.05).

\section*{Author contributions: CRediT}
Marta A. Chabowska: Conceptualization, Data curation, Formal analysis, Investigation, Methodology, Validation, Visualization, Writing - original draft, Writing - review and editing;
Hristina Popova: Data curation, Formal analysis, Investigation, Validation, Visualization, Writing - original draft, Writing - review and editing;
Magdalena A. Za{\l}uska - Kotur: Conceptualization, Formal analysis, Investigation, Methodology, Validation, Writing - original draft, Writing - review and editing.

\newpage

\section {Supplementary Material}
\renewcommand{\thefigure}{\textbf{S\arabic{figure}}}
\setcounter{figure}{0}

In this supplement we provide the additional figures.

\begin{figure*}[hbt]
 \centering
\includegraphics[width=0.9\textwidth]{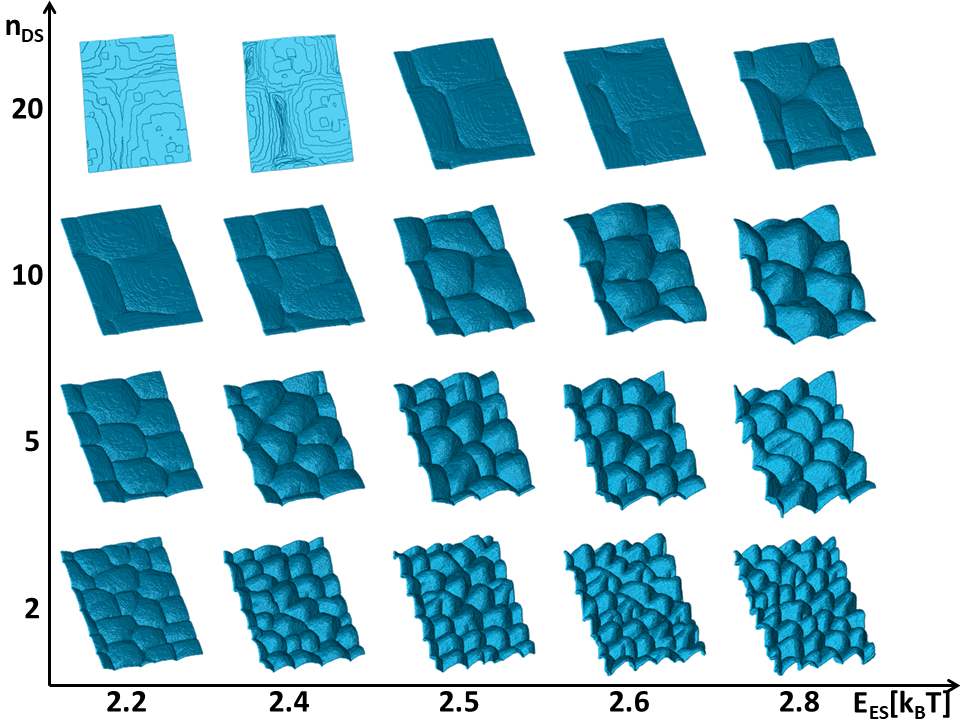}
\caption{ Diagram of different surface patterns presented as a function of number of diffusional jumps $n_{DS}$ dependent on the height of ES barrier $E_{ES}$ for flat surface, $E_V = 1.0 k_B T$ and initial particle concentration $c_0=0.01$.  System size $300$ x $400$ and number of simulation steps $t=10^6$.}
\label{S1-fig}
\end{figure*}

Figure~\ref{S1-fig} presents a diagram of patterns grown on flat surface as a function of number of diffusional jumps $n_{DS}$ with a more detailed examination of the morphological transition controlled by the height of the Ehrlich-Schwoebel (ES) barrier from $E_{ES} = 2k_B T$ toward $E_{ES} = 3k_B T$.

\begin{figure*}[hbt]
 \centering
 \includegraphics[width=1.0\textwidth]{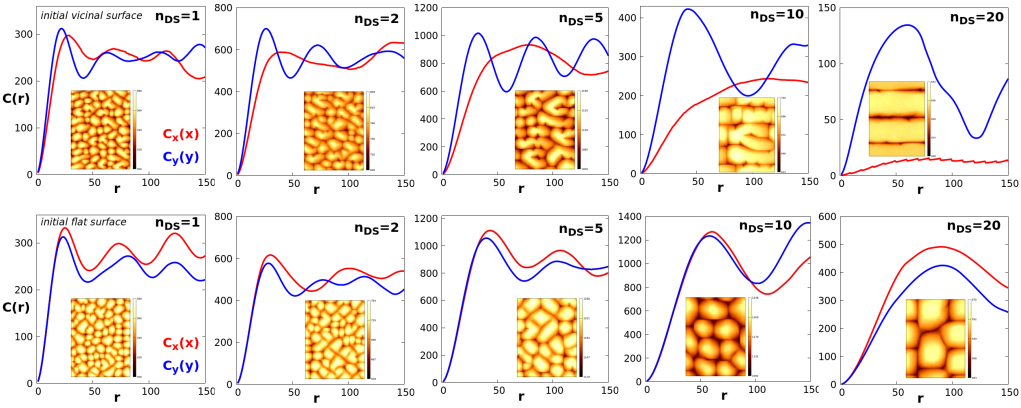}
\caption{Comparison of the correlation functions calculated along $x$- and $y$-axis for 3D structures grown on vicinal and flat surfaces (top and bottom panel, respectively) for different diffusion rates $n_{DS}$ (shown in the legend). Other system parameters are: $c_0 = 0.01, l_0 = 10$ (for vicinal surface), $E_V = 1 k_BT, E_{ES} = 3 k_BT$, simulation time $10^6$, and system size $300$ x $400$. The corresponding surface morphologies are also presented.}
\label{fig_comparing_3D_structures_on_flat_and_vicinal_surfaces}
\end{figure*}

Figure~\ref{fig_comparing_3D_structures_on_flat_and_vicinal_surfaces} provides a comparison of the correlation functions $C(r)$ calculated along $x$- and $y$-axis separately for three-dimensional (3D) structures grown at different diffusion rates on vicinal and flat surfaces, respectively.

\begin{figure*}[hbt]
 \centering
 \includegraphics[width=0.6\textwidth]{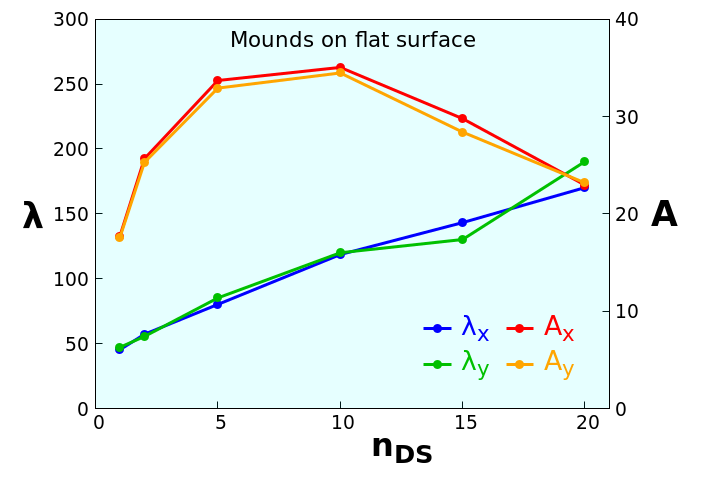}
\caption{Influence of diffusion rate $n_{DS}$ on the main morphological characteristics $\lambda$ and $A$ calculated along $x$- and $y$-axis separately for mounds grown on flat surface. Other system parameters are: $c_0 = 0.01, E_V = 1 k_BT, E_{ES} = 3 k_BT$, simulation time $10^6$, and system size $300$ x $400$.}
\label{fig_lambda_amplitude_nds_influence_flat}
\end{figure*}

Analyzing the mounded morphologies on flat surface, Figure~\ref{fig_lambda_amplitude_nds_influence_flat} demonstrates the influence of the diffusion rate $n_{DS}$ on the main morphological characteristics of mounds, (lateral) wavelength $\lambda$ and (vertical) amplitude $A$, which are calculated from correlation functions along $x$- and $y$-axis separately. Typical of isotropic mounds, each of the characteristics has the same behavior in both directions. However, for large mounds (with lateral size $\lambda$ larger than the half of the system size in particular direction) the finite-size effect of the system must be taken into account causing a small deviation in the results.

\begin{figure*}[hbt]
 \centering
 \includegraphics[width=0.38\textwidth]{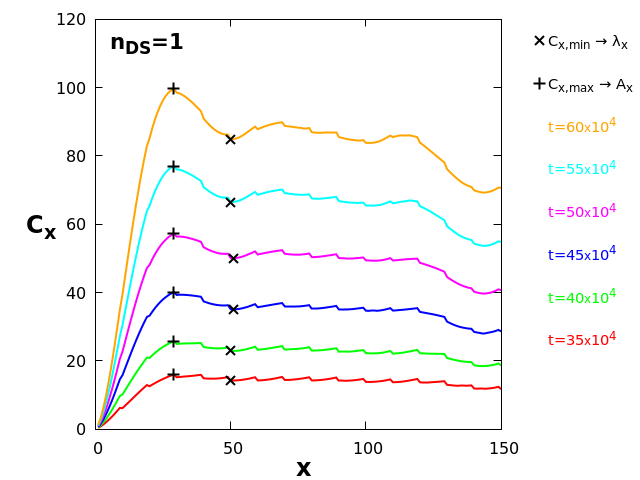}
\hspace{-1.5 cm}
\includegraphics[width=0.38\textwidth]{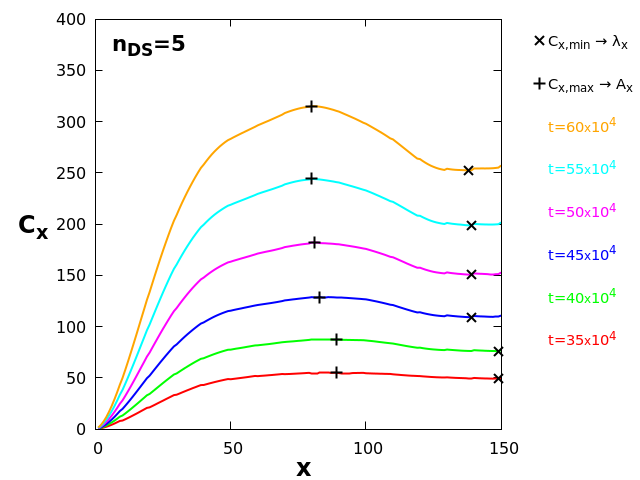}
\hspace{-1.5 cm}
\includegraphics[width=0.38\textwidth]{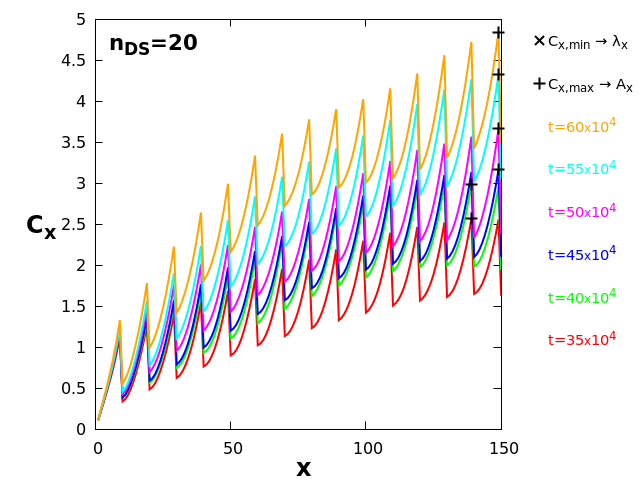}
\includegraphics[width=0.38\textwidth]{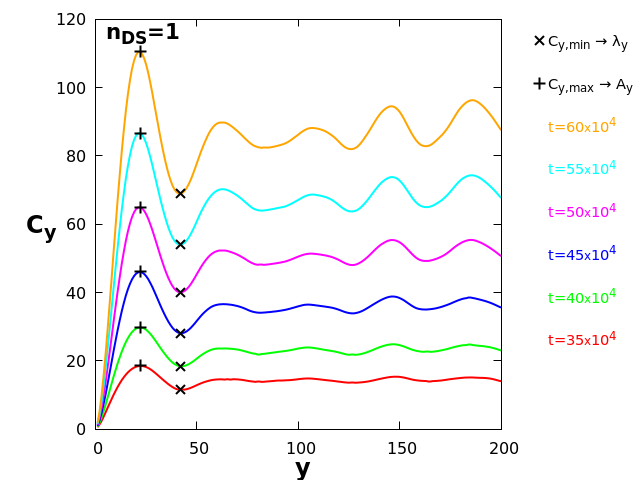}
\hspace{-1.5 cm}
\includegraphics[width=0.38\textwidth]{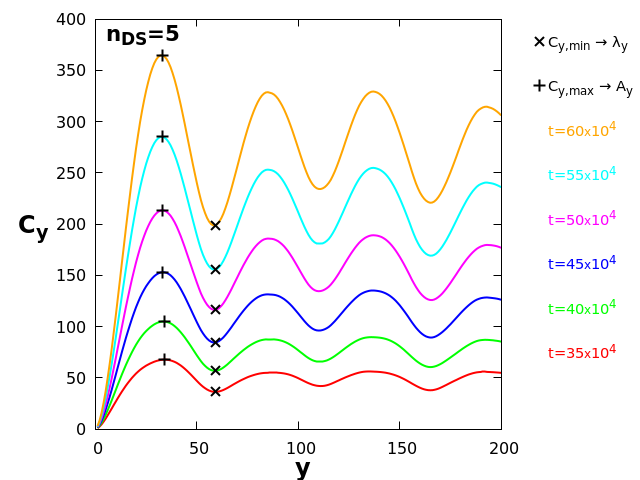}
\hspace{-1.5 cm}
\includegraphics[width=0.38\textwidth]{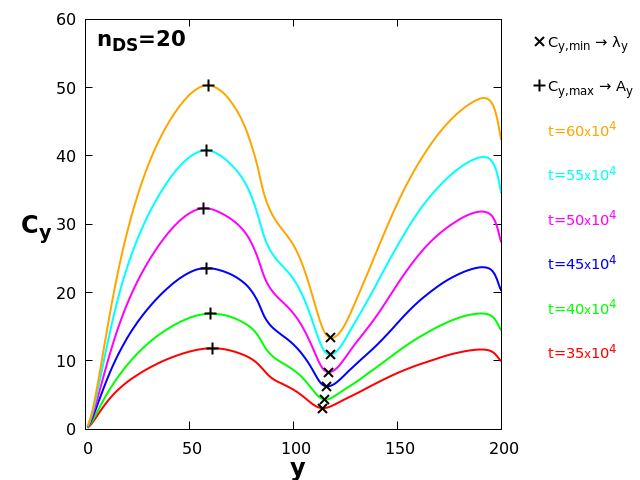}
\caption{Correlation functions $C_x(x)$ (top panel) and $C_y(y)$ (bottom panel) for vicinal surface presented for different time moments $t$ as indicated in the legend, demonstrating the evolution of mounded structure (in the left, for $n_{DS}=1$), structure consisting simultaneously of mounds and meanders (in the middle, for $n_{DS}=5$) and meandered structure (in the right, for $n_{DS}=20$). Other system parameters are: $c_0 = 0.01, l_0 = 10, E_V = 1 k_BT, E_{ES} = 3 k_BT$. System size is $300$ x $400$. Positions of the first minimum and the first maximum of the correlation functions, corresponding to the characteristic lengths $\lambda$ and $A$, are marked on each curve to easily track their evolution.}
\label{fig_new-cor-time-nds}
\end{figure*}

\begin{figure*}[hbt]
 \centering
 \includegraphics[width=0.49\textwidth]{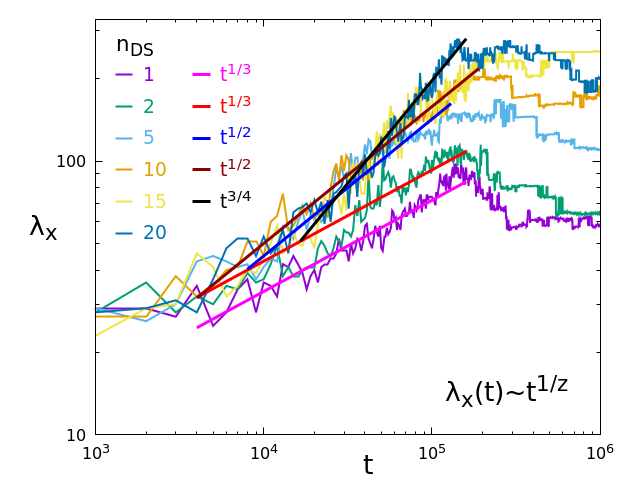}
 \includegraphics[width=0.49\textwidth]{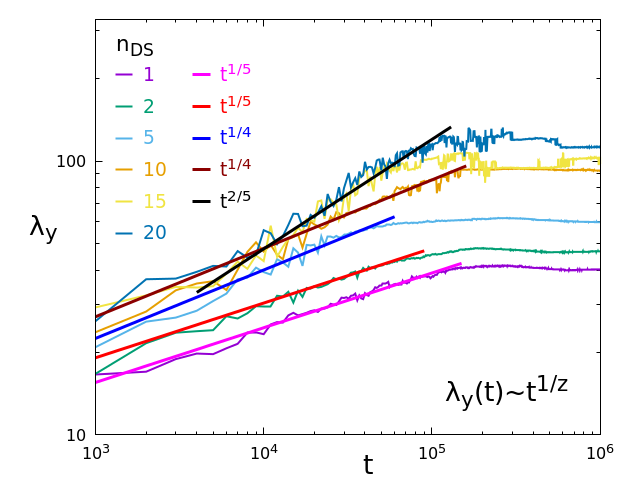}
 \includegraphics[width=0.49\textwidth]{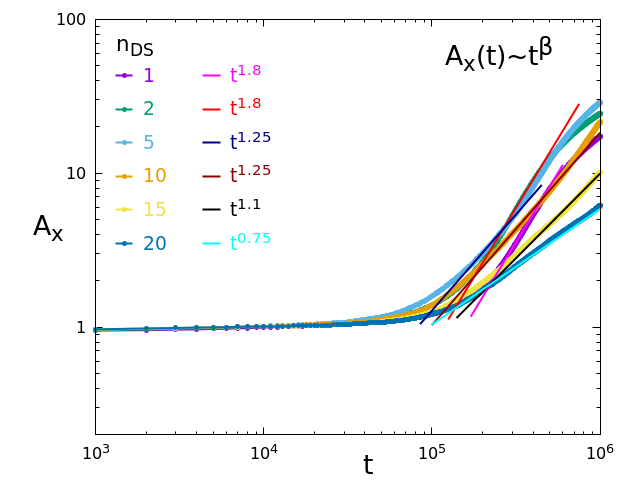}
 \includegraphics[width=0.49\textwidth]{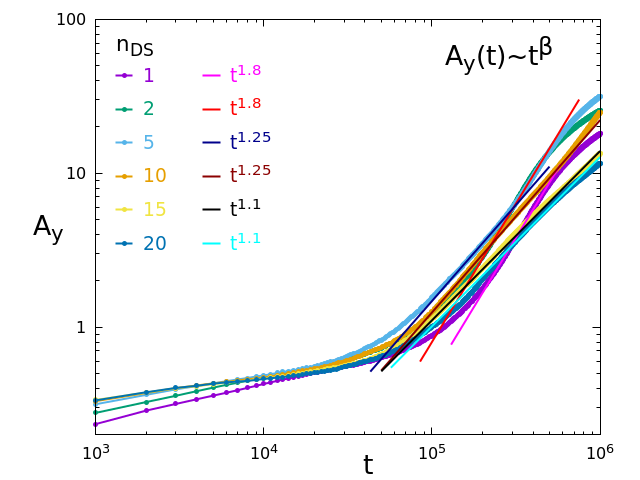}
\caption{Time scaling of the main characteristic lengths - the wavelengths $\lambda_x$ and $\lambda_y$ (top panel) and the amplitudes $A_x$ and $A_y$ (bottom panel) for structures grown at different diffusion rate $n_{DS}$ on vicinal surface. Time-scaling relations of characteristic lengths and their corresponding scaling exponents are shown in the legends. System parameters are: $c_0 = 0.01, l_0 = 10, E_V = 1 k_BT, E_{ES} = 3 k_BT$. System size is $300$ x $400$.}
\label{fig_new-time-scaling-nds}
\end{figure*}

Figures~\ref{fig_new-cor-time-nds} and \ref{fig_new-time-scaling-nds} present the results obtained for a vicinal surface. Figure~\ref{fig_new-cor-time-nds} shows the correlation functions $C_x(x)$ and $C_y(y)$ for different time moments $t$ describing the evolution of the surface morphology and the corresponding characteristic lengths along both directions. The separate graphs represent different values of $n_{DS}$ demonstrating the evolution of structures consisting of regular mounds, regular meanders and a mix of both.

Figure~\ref{fig_new-time-scaling-nds} presents the time scaling of the wavelengths $\lambda$ and the amplitudes $A$ in both directions for structures grown at different diffusion rate $n_{DS}$ on vicinal surface. The dynamics of the crystal surface is characterized by scaling exponents describing the increase of the wavelength $\lambda \sim t^{1/z}$ and the increase of the amplitude $A \sim t^{\beta}$ during the growth process. Note that both scaling exponents, $1/z$ and $\beta$, are measured in different time intervals corresponding to the steadily increase of the particular characteristic length. Initially both wavelengths, $\lambda_x$ and $\lambda_y$, exhibit fast increase in time, while the amplitude stays almost constant and low, which implies mainly lateral growth. Later in time, the amplitude starts to grow drastically, indicating vertical growth and the increase in surface roughness, in contrast to the wavelengths in both directions which remain almost the same. Therefore, coarsening (coalescing) of mounds or meanders is observed in the initial stage of the growth process, and further surface roughening appears in the late stage. Although the scaling dynamics slightly differs in $x$ and $y$ direction, the transition from mounds to meanders by increasing of $n_{DS}$ is reflected in the change of all scaling exponents, such that the dynamic (coarsening) exponent $1/z$ increases and the growth (roughness) exponent $\beta$ decreases simultaneously in both directions.

\end{document}